\documentclass[aps,pra,9pt,twocolumn,showpacs,superscriptaddress]{revtex4-2}
\usepackage{physics}							
\usepackage{latexsym}
\usepackage{amssymb}
\usepackage{graphics,epstopdf}
\usepackage{newlfont}
\usepackage{amsfonts}
\usepackage{epsfig}
\usepackage[colorlinks=true, citecolor=blue, urlcolor=blue]{hyperref}
\usepackage{amsmath}
\usepackage{graphicx}
\usepackage{dcolumn}% Align table columns on decimal point
\usepackage{bm}% bold math
\usepackage{color}
\usepackage{longtable}
\usepackage{amsthm}
\usepackage{times}
\newtheorem{theorem}{Theorem}
\newtheorem*{theorem*}{Theorem}

\newtheorem{proposition}{Proposition}

\begin{document}

\title{Quantum Speed Limit From Tighter Uncertainty Relation}

\author{Shrobona Bagchi}
\email{shrobona@kist.re.kr}
\affiliation{Center for Quantum Information, Korea Institute of Science and Technology, Seoul, 02792, Korea}
\author{Abhay Srivastav}
\email{abhaysrivastav@hri.res.in}

\author{Arun Kumar Pati}
\email{akpati@hri.res.in}
\affiliation{\(\)Harish-Chandra Research Institute,  A CI of Homi Bhabha National Institute, Chhatnag Road, Jhunsi, Prayagraj  211019, Uttar Pradesh, India
}

\begin{abstract}
%Time energy uncertainty relation is one of the cornerstone in quantum physics. One of the interpretations of the time energy uncertainty relation is the quantum speed limit. 
The quantum speed limit provides a fundamental bound on how fast a quantum system can evolve between the initial and the final states under any physical operation. The celebrated Mandelstam-Tamm (MT) bound has been widely studied for various quantum systems undergoing unitary time evolution. Here, we prove a new quantum speed limit using the tighter uncertainty relations for pure quantum systems undergoing arbitrary unitary evolution. We also derive a tighter uncertainty relation for mixed quantum states and then derive a new quantum speed limit for mixed quantum states from it such that it reduces to that of the pure quantum states derived from tighter uncertainty relations. We show that the MT bound is a special case of the tighter quantum speed limit derived here. We also show that this bound can be improved when optimized over many different sets of basis vectors. We illustrate the tighter speed limit for pure states with examples using random Hamiltonians and show that the new quantum speed limit outperforms the MT bound.
\end{abstract}

\maketitle

\section {Introduction }\label{intro}
The uncertainty principle and the uncertainty relations are of central importance in quantum mechanics. The uncertainty relations have helped us to reveal the behavior of the microscopic world in many different ways. At first the uncertainty principle was discovered by Heisenberg who heuristically provided a lower bound on the product of the error and the disturbance for two canonically conjugate quantum mechanical observables \cite{Heisenberg1927}. On the other hand, the uncertainty relations are capable of capturing the intrinsic restrictions in preparation of quantum systems, which are termed as the preparation uncertainty relations \cite{Robertson1929}. This interpretation was quite fruitful for the uncertainty relations like position-momentum, angular position-angular momentum uncertainty relations etc. However, the energy-time uncertainty relation \cite{Aharonov1961,Aharonov2002} is different from the above stated uncertainty relations because time is not treated as an operator in quantum mechanics but as a classical parameter with no inherent quantum uncertainty in it~\cite{Busch2008}. The uncertainty relation for two arbitrary quantum-mechanical observables formulated by Robertson is essentially a preparation uncertainty relation and expresses the impossibility of joint sharp preparation of any two incompatible observables. However, the Robertson uncertainty relation does not completely express the incompatible nature of two non-commuting observables in terms of uncertainty quantification. To capture the notion of incompatibility more efficiently a stronger form of the uncertainty relation based on the sum of variances was derived in Ref.~\cite{Pati2014}. In addition, tighter uncertainty relations and reverse uncertainty relations have also been proved which go beyond the Robertson-Schr{\"o}dinger uncertainty relations \cite{Mondal2017}. The stronger uncertainty relations and the reverse uncertainty relations have been experimentally tested using photonic set-ups \cite{xiao2020}. However, time, not being a quantum observable, energy-time uncertainty relation lacked a good interpretation as such like for those of the other quantum mechanical observables such as position and momentum.

It was shown by Mandelstam and Tamm that the correct interpretation of the energy-time uncertainty relation is as a bound on the evolution time of a quantum system, now known as the MT bound \cite{Mandelstam1945}. 
%In the existing literature, there are several other approaches to obtain quantum speed limits \cite{10,11,12,13}. 
Subsequently, Margolus and Levitin derived a new bound on the evolution time based on the expectation value of the Hamiltonian \cite{Margolus1998}. The quantum speed limit bounds have since been studied extensively for closed \cite{Anandan1990, Levitin2009, Gislason1956, Eberly1973, Bauer1978, Bhattacharyya1983, Leubner1985,  Vaidman1992, Uhlmann1992, Uffink1993, Pfeifer1995,  Horesh1998, AKPati1999, Soderholm1999,  Andrecut2004, Gray2005, Luo2005,  Zielinski2006,  Andrews2007, Yurtsever2010, Fu2010, Zwierz2012, Poggi2013,Kupferman2008, Jones2010, Chau2010, S.Deffner2013, Fung2014, Andersson2014, D.Mondal2016, Mondal2016, S.Deffner2017, Campaioli2018,Giovannetti2004, Batle2005, Borras2006, Zander2007,Ness2022, Pandey22, thakuria2022} as well as for open system dynamics \cite{Deffner2013, Campo2013, Taddei2013, Fung2013, Pires2016, S.Deffner2020,Jing2016,Pintos2021,Mohan,Mohan21,Pandey2022}. Recently the notion of quantum speed limit has been generalised for arbitrary evolution \cite{abhay2022}, unitary operator flows \cite{Carabba2022}, change of basis \cite{naseri2022}, and in arbitrary phase spaces \cite{meng2022}. 
%The quantum speed limit bounds have also been extended to the case of mixed quantum states undergoing unitary evolution \cite{14,15,16,17,18,19,20,21,22,23,24,25,26,27,28}, entangled quantum states \cite{29,30,31} and also open quantum system dynamics \cite{32,33,34}. 

The notion of quantum speed limit is not only of fundamental importance, but also has practical applications in quantum information and quantum technology. The quantum speed limit bounds have proven to be very useful in quantifying the maximal rate of quantum entropy production \cite{deffner2010,das2018}, the maximal rate of quantum information processing \cite{bekenstein1981,Mohan}, quantum computation \cite{lloyd2000, lloyd2002, AGN12} in optimal control theory \cite{Caneva2009, Campbell2017}, quantum thermometry \cite{Campbell2018}, quantum thermodynamics~\cite{Mukhopadhyay2018} etc. These explorations motivate us to find better quantum speed limit bounds that can go beyond the existing bounds in the literature. In this paper, we use the tighter uncertainty relation \cite{Mondal2017} to derive a tighter form of quantum speed limit for pure as well as mixed states undergoing unitary evolution. We show that the new bound provides a tighter expression of quantum speed limit compared to the MT bound. This bound can also be optimized over many orthonormal basis vector sets, as in the case of tighter uncertainty relations. We then find various examples for pure states that shows the better performance of our bound over the MT bound and the bound in Ref.~\cite{Campaioli2018}. 

The present article is organised as follows. In Section \ref{background}, we give the background needed for our result that includes a brief review of quantum speed limit and the tighter uncertainty relations for pure quantum states. In section \ref{TUR for mixed states}, we provide the derivation of tighter uncertainty relation for the case of mixed quantum states. In section \ref{TQSL for pure states} and \ref{TQSL for mixed states}, we derive the tighter quantum speed limit for pure and mixed quantum states, respectively, and show that the MT bound is a special case. In Section \ref{examples}, we analyse the obtained bound for some numerical examples using random Hamiltonians chosen from the Gaussian Unitary Ensemble and find that the new bound surpasses the MT bound for the case of pure quantum states. We also show the better performance of our bound for an interacting quantum system. Finally, in Section \ref{conclusion}, we conclude and point out future directions.

\section {Background }\label{background}

\subsection {Standard Quantum Speed Limits}
Quantum speed limit (QSL) is a fundamental limitation on the speed of the evolution of a quantum system imposed by the laws of quantum mechanics. Historically, Mandelstam and Tamm derived the first expression of the quantum speed limit time as $\tau_{QSL} = \frac{\pi}{2\Delta H}$, where $\Delta H$ is the standard deviation of the Hamiltonian $H$ driving the quantum system, and where the initial and the final states are orthogonal. As a physical interpretation of their bound, they also argued that $\tau_{QSL}$ quantifies the life time of quantum states, which has found importance in the foundations of quantum mechanics as well as in other applications. Their interpretation was further solidified by Margolus and Levitin, who derived an alternative expression for $\tau_{QSL}$ in terms of the expectation value of the Hamiltonian as $\tau_{QSL} = \frac{\pi}{2\langle H\rangle}$. Eventually, it was also shown that the combined bound,
\begin{align}
\tau_{QSL}=\max\left\{\frac{\pi\hbar}{2\Delta H},\frac{\pi\hbar}{2\langle H\rangle}\right\}
\end{align}
is tight for the evolution of the system between two orthogonal states \cite{Levitin2009}. The QSL bound can be generalised for the evolution between two non-orthogonal states using the Fubini-Study metric on the projective Hilbert space given by
\begin{eqnarray}
 \label{FSM}
 ds^2 = 4 [\langle \dot{\Psi} (t)|\dot{\Psi}(t)\rangle-(i\langle \Psi (t)|\dot{\Psi}(t)\rangle)^2 ]dt^2,
 \end{eqnarray}
and the Schr\"odinger equation for the unitary evolution of a quantum state. The fact that the total distance travelled by a quantum state in the projective Hilbert space is always greater than or equal to the shortest distance connecting the initial and the final points, i.e., the geodesic $s_0$ implies 
\begin{eqnarray}
\tau \geq \tau_{QSL}= \frac{\hbar s_0}{ 2 {\Delta{H}}},
\end{eqnarray}
where $\tau$ is the actual time of evolution and $s_0(t)=2\cos^{-1}|\langle\Psi(t)|\Psi(0)\rangle|$. $\tau_{QSL}$ is the celebrated MT bound and gives the minimum time required for a quantum system to evolve between any two states unitarily. In \cite{Campaioli2018} another bound tighter than the MT bound was derived for the speed of unitary evolution. This bound for time independent Hamiltonian and pure quantum states is given as follows
\begin{align}
\tau\geq\tau_2=\sqrt{1-\frac{1}{N}}\frac{\cos^{-1}\left(\frac{|\innerproduct{\Psi(0)}{\Psi(\tau)}|^2-\frac{1}{N}}{1-\frac{1}{N}}\right)}{\sqrt{2}\Delta H}, \label{better bound}
\end{align}
where $N$ is the dimension of the quantum system undergoing unitary evolution due to the time independent Hamiltonian $H$. We mention this bound since this bound does not reduce to the MT bound in general.

\subsection{Tighter Uncertainty Relations for pure quantum states}

Uncertainty relations hold an important place in the foundations of quantum mechanics. A quantitative formulation of the Heisenberg uncertainty principle was given by Robertson. This is also known as the Robertson-Schr\"odinger uncertainty relation. For any two generally non-commuting operators A and B, the Robertson-Schr\"odinger uncertainty relation for the state of the system $|\Psi\rangle$ is given by the following inequality:
\begin{align}
\Delta A^2\Delta B^2\geq |\frac{1}{2}\langle[A,B]\rangle|^2+|\frac{1}{2}\langle\{A,B\}\rangle-\langle A\rangle\langle B\rangle|^2,
\end{align}
where the averages and the variances are defined over the state of the system $|\Psi\rangle$. This relation is a direct consequence of the Cauchy-Schwarz inequality. However, this
uncertainty bound is not optimal. There have been several attempts to tighten the bound, for example see \cite{Pati2014,Mondal2017}. Here, we state a tighter bound which can be expressed as
\begin{align}
\Delta A^2\Delta B^2\geq \max_{\{|\psi_n\rangle\}}\frac{1}{2}|\langle[\bar{A},\bar{B}^{\psi}_n]_{\Psi}\rangle|^2+|\frac{1}{2}\langle\{\bar{A},\bar{B}^{\psi}_n\}_{\Psi}\rangle|^2,
\end{align}
where $\bar{B}^{\psi}_n=|\psi_n\rangle\langle\psi_n|(B-\langle B\rangle)$, $\bar{A}=A-\langle A\rangle $ and $\{|\psi_n\rangle\}$ is the eigenbasis of any observable other than $A$ and $B$ \cite{Mondal2017}. This uncertainty relation was proved to be tighter than Robertson-Schr\"odinger uncertainty relation and even outperforms the stronger uncertainty relations by Maccone-Pati \cite{Pati2014}, in some cases. We will use this tighter uncertainty relation for deriving a tighter quantum speed limit bound in the following sections. The derivation of this type of tighter uncertainty relation for mixed states is given in the next section.

\section{Tighter Uncertainty Relations for mixed quantum states} \label{TUR for mixed states}
\theorem{The tighter uncertainty relation for two non-commuting operators A and B for the mixed quantum state $\rho$ is given by the following inequality
\begin{align}\nonumber 
\Delta A\Delta B\geq \sum_n\sqrt{|\mathrm{Tr}(\bar{A}\rho\bar{A}\bar{B}^{\psi}_n\rho\bar{B}^{\psi}_n)|} \geq |\mathrm{Tr}(\bar{A}\rho\bar{B})|,
\end{align}
where 
\begin{align}\nonumber
\bar{A}=A-\mathrm{Tr}(\rho A)\mathbb{I},~\bar{B}=B-\mathrm{Tr}(\rho B)\mathbb{I},~
\bar{B}^{\psi}_n=|\psi_n\rangle\langle\psi_n|\bar{B},\nonumber
\end{align}
and $\{|\psi_n\rangle\}$ form a complete orthonormal basis.}
\proof{For proving the tighter uncertainty relation for mixed states, we define the following quantities:
\begin{eqnarray}
f&=&\bar{A}\rho \bar{A}=\sum_{m,n}\alpha_{m,n}|\psi_m\rangle\langle\psi_n|, \nonumber\\
g&=&\bar{B}\rho \bar{B}=\sum_{i,j}\beta_{i,j}|\psi_i\rangle\langle\psi_j|,   \label{fgdefinition}
\end{eqnarray}
where $\bar{A}$ and $\bar{B}$ are as defined above, and $\alpha_{m,n}$ and $\beta_{i,j}$ are complex numbers.
%The above equation holds because we know that $\{|\Psi\rangle_i\}$ form a complete basis in the Hilbert space and therefore we can express any operator in that basis with complex coefficients in general. 
We will now prove that the operators $f$ and $g$ are Hermitian with non-negative eigenvalues and therefore are positive operators. We prove it explicitly for $f$ and the same follows for $g$.
Let us define an operator $F$ as follows
\begin{align}
F=\bar{A}\sqrt{\rho},
\end{align}
where we have taken the positive square root of $\rho$ without any loss of generality.
Using the above equation then we get 
\begin{align}
FF^\dagger=(\bar{A}\sqrt{\rho})(\bar{A}\sqrt{\rho})^\dagger=(\bar{A}\sqrt{\rho})(\sqrt{\rho}^\dagger\bar{A}^\dagger).
\end{align}
This implies that
\begin{align}
FF^\dagger=(\bar{A}\sqrt{\rho})(\sqrt{\rho}\bar{A})=\bar{A}\rho\bar{A}=f.\label{ffdagger}
\end{align}
%The above equation holds because we know that both $\sqrt{\rho}$ and $\bar{A}$ are Hermitian operators. Thus we have proved that $f=FF^\dagger$.
where we have used the fact that $\bar{A}$ $\sqrt{\rho}$ are Hermitian operators. The hermiticity of $\sqrt{\rho}$ can be proved as follows. Let $U$ be the unitary that diagonalises $\rho$ as follows
%the basis in which $\rho$ is diagonal be $\{|\Phi_i\rangle\}$. Therefore we have the following 
%\begin{align}
%\rho_d=\sum_i e_i|\Phi_i\rangle\langle\Phi_i|
%\Rightarrow \sqrt{\rho_d}=\sum_i \sqrt{e_i}|\Phi_i\rangle\langle\Phi_i|
%\end{align}
%Now we know that $\rho$ is a positive semidefinite Hermitian operator. As a result, $e_i$ are all real and positive semidefinite. Thus, $\sqrt{e_i}$ are also real. Now we again note the following by using the fact that the density matrices are diagonalizable by unitary operators as follows
\begin{align}
\rho=U\rho_d U^\dagger
\Rightarrow \sqrt{\rho}=U\sqrt{\rho_d} U^\dagger,
\end{align}
%where we have used a valid method to construct $\sqrt{\rho}$ from $\rho$. 
Using the above equation then we see that
%Therefore, if we take the adjoint of $\sqrt{\rho}$, we get the following equation 
\begin{eqnarray}
(\sqrt{\rho})^\dagger&=&(U\sqrt{\rho_d} U^\dagger)^\dagger 
=U\sqrt{\rho_d} U^\dagger\nonumber\\&=&\sqrt{\rho}.
\end{eqnarray}
Therefore, $\sqrt{\rho}$ is a Hermitian operator. Now, using the hermiticity property of $F$ and Eq.\eqref{ffdagger} we see that $f$ is also Hermitian as 
\begin{align}
f^\dagger=(FF^\dagger)^\dagger=(F^\dagger)^\dagger F^\dagger=FF^\dagger=f.
\end{align}
Now, we analyze the eigenvalues of the operator $f=FF^\dagger$. For this let $F$ be diagonalizable in the basis $\{|k\rangle\}$ with eigenvalues $F_k$, such that we have the following equation 
\begin{align}
F=\sum_k F_k|k\rangle\langle k|.
\end{align}
Then we have the eigenvalues of the operator $f$ from the following equation 
\begin{align}
f=FF^\dagger=
%\sum_k F_k|k\rangle\langle k|\sum_i F_i^*|i\rangle\langle i|\\ \nonumber =\sum_{ik} F_kF_i^*|k\rangle\langle k|i\rangle\langle i|=\sum_{ik} F_kF_i^*|k\rangle\langle i|\delta_{ik}\\ \nonumber=\sum_{i} F_iF_i^*|i\rangle\langle i|=
\sum_{i} |F_i|^2|i\rangle\langle i|.
\end{align}
Thus, from the above equation we see that $f$ is diagonalized in the same basis as $F$ and all is eigenvalues $|F_i|^2$ are real and positive semidefinite. Now, we know that if all of the eigenvalues of a self adjoint or Hermitian operator are positive semidefinite, then that operator is positive semidefinite. Now from the above observations, we note the following properties about the nature of $\alpha_{m,n}$ and $\beta_{i,j}$. Since $f$ is a positive operator, therefore we have $\langle x,fx\rangle >0$ for all $x\neq 0$. 
%As a definition of positive operator we have that given a Hilbert space  $H$ and  $A\in L(H)$,  $A$ is said to be a positive operator if  $\langle x|A|x\rangle\geq 0$ for every  $x \in H$. 
From this definition, we have 
$\langle\psi_n|f|\psi_n\rangle > 0 $ which implies that $\alpha_{n,n}>0$ $\forall~n$.
Similarly, $\beta_{m,m}>0$ $\forall~m$ as well. Keeping in mind that these properties hold, we now move on to prove the tighter uncertainty relation for mixed quantum states. From our definitions we get the following
\begin{eqnarray}
\Delta A^2&=&\mathrm{Tr}(A^2\rho)-\mathrm{Tr}(A\rho)^2=\mathrm{Tr}(\bar{A}^2\rho)=\mathrm{Tr}(f) \nonumber\\ \Delta B^2&=&\mathrm{Tr}(B^2\rho)-\mathrm{Tr}(B\rho)^2=\mathrm{Tr}(\bar{B}^2\rho)=\mathrm{Tr}(g).\nonumber
\end{eqnarray}
Therefore, we have 
$\Delta A\Delta B=\sqrt{\mathrm{Tr}(f)\mathrm{Tr}(g)}$.
Using the definition of $f$ and $g$ we have 
\begin{align}
\mathrm{Tr}(f)\mathrm{Tr}(g)= \sum_{m,n}\alpha_{n,n}\beta_{m,m}.
\end{align}
Now we know from the structure of $\{f,g\}$ that $\alpha_{n,n}$ and $\beta_{m,m}$ are real positive numbers in general. Therefore, using the Cauchy-Schwarz inequality for two real positive vectors $\{|\alpha_1|,|\alpha_2|...,|\alpha_n|\}$ and $\{|\beta_1|,|\beta_2|...,|\beta_n|\}$ we have the following inequality 
\begin{align}\nonumber
\mathrm{Tr}(f)\mathrm{Tr}(g)=\sum_{m,n}\alpha_{n,n}\beta_{m,m}\geq \left(\sum_{n}|\sqrt{\alpha_{n,n}}\sqrt{\beta_{n,n}}|\right)^2. 
%=(\sum_{n}|\sqrt{\alpha_{n,n}^*}\sqrt{\beta_{n,n}}|)^2
\end{align}
Now, using Eq.\eqref{fgdefinition} we get the following inequality
\begin{eqnarray}\nonumber
\mathrm{Tr}(f)\mathrm{Tr}(g)&\geq&
\left(\sum_n\sqrt{|\langle\psi_n|\bar{A}\rho\bar{A}|\psi_n\rangle\langle\psi_n|\bar{B}\rho\bar{B}|\psi_n\rangle|}\right)^2\\ \nonumber 
&=&\left(\sum_n\sqrt{|\mathrm{Tr}(\bar{A}\rho\bar{A}|\psi_n\rangle\langle\psi_n|\bar{B}\rho\bar{B}|\psi_n\rangle\langle\psi_n|)|}\right)^2 \\ \nonumber 
&=&\left(\sum_n\sqrt{|\mathrm{Tr}(\bar{A}\rho\bar{A}\bar{B}^{\psi}_n\rho\bar{B}^{\psi}_n|)|}\right)^2,
\end{eqnarray}
where we have defined the operator $\bar{B}^{\psi}_n=|\psi_n\rangle\langle\psi_n|\bar{B}$. Therefore, we get the following as the mixed state version of the tighter uncertainty relation
\begin{align}
\Delta A\Delta B\geq \sum_n\sqrt{|\mathrm{Tr}(\bar{A}\rho\bar{A}\bar{B}^{\psi}_n\rho\bar{B}^{\psi}_n)|}.  \label{tightuncrelation}
\end{align}
%It is straightforward to see that the above relation reduces to that of the pure states version when we take $\rho=|\Psi\rangle\langle\Psi|$. 
The above equation holds true whichever way we define $\bar{B}^{\psi}_n$, i.e., either as $|\psi_n\rangle\langle\psi_n|\bar{B}$, or as $\bar{B}|\psi_n\rangle\langle\psi_n|$, which is straightforward to deduce from the above equations.
Let us again consider the bound given in Eq.\eqref{tightuncrelation} 
\begin{eqnarray}
&&\sum_n\sqrt{|\mathrm{Tr}(\bar{A}\rho\bar{A}\bar{B}^{\psi}_n\rho\bar{B}^{\psi}_n)|}\nonumber\\
&=&\sum_n\sqrt{|\mathrm{Tr}(\bar{A}\rho\bar{A}|\psi_n\rangle\langle\psi_n|\bar{B}\rho\bar{B}|\psi_n\rangle\langle\psi_n|}\nonumber\\
&=&\sum_n\sqrt{|\langle\psi_n|\bar{A}\rho\bar{A}|\psi_n\rangle\langle\psi_n|\bar{B}\rho\bar{B}|\psi_n\rangle|}\nonumber\\
&=&\sum_n\sqrt{|\mathrm{Tr}(\bar{A}\rho\bar{A}|\psi_n\rangle\langle\psi_n|)||\mathrm{Tr}((\bar{B}\rho\bar{B}|\psi_n\rangle\langle\psi_n|}.\nonumber
\end{eqnarray}
Now using Cauchy-Schwarz inequality for complex matrices, we get
\begin{eqnarray}
&&\sum_n\sqrt{|\mathrm{Tr}(\bar{A}\rho\bar{A}|\psi_n\rangle\langle\psi_n|)||\mathrm{Tr}((\bar{B}\rho\bar{B}|\psi_n\rangle\langle\psi_n|}\nonumber\\
&\geq& \sum_n\sqrt{|\mathrm{Tr}(\langle\psi_n|\bar{A}\sqrt{\rho}\sqrt{\rho}\bar{B}|\psi_n\rangle)|^2}\nonumber\\ 
&=&\sum_n|\mathrm{Tr}(\langle\psi_n|\bar{A}\rho\bar{B}|\psi_n\rangle)|
\geq |\sum_n\mathrm{Tr}(\langle\psi_n|\bar{A}\rho\bar{B}|\psi_n\rangle)|\nonumber\\
&=& |\mathrm{Tr}(\sum_n\bar{A}\rho\bar{B}|\psi_n\rangle\langle\psi_n|)|=|\mathrm{Tr}(\bar{A}\rho\bar{B})|.
\end{eqnarray}
Therefore, we have proved the following for our tighter uncertainty relation for mixed quantum states $\rho$.
\begin{align}
\Delta A\Delta B\geq \sum_n\sqrt{|\mathrm{Tr}(\bar{A}\rho\bar{A}\bar{B}^{\psi}_n\rho\bar{B}^{\psi}_n)|} \geq |\mathrm{Tr}(\bar{A}\rho\bar{B})|. \label{tighterthanRobertSch}
\end{align}
%where we have defined $\bar{B}^{\Psi}_n=|\Psi_n\rangle\langle\Psi_n|\bar{B}$.
We know that the term on the right hand side gives us the bound given by the Robertson-Schrodinger uncertainty relation. As a result therefore we have shown that the new uncertainty relation derived here for mixed quantum states outperforms the Robertson-Schrodinger uncertainty relation for mixed quantum states. We will now show that the above uncertainty relation reduces to the tighter uncertainty relation for that of the pure states when we have $\rho=|\Psi\rangle\langle\Psi|$. Using this $\rho$ in Eq.\eqref{tighterthanRobertSch} we get 
\begin{eqnarray}\nonumber 
\Delta A\Delta B&\geq& \sum_n\sqrt{|\mathrm{Tr}(\bar{A}|\Psi\rangle\langle\Psi|\bar{A}\bar{B}^{\psi}_n|\Psi\rangle\langle\Psi|\bar{B}^{\psi}_n|}\\ \nonumber  
&\geq& |\mathrm{Tr}(\bar{A}|\Psi\rangle\langle\Psi|\bar{B})|.
\end{eqnarray}
Simplifying the above equation we get
%\begin{align}\nonumber 
%\Delta A\Delta B\geq \sum_n\sqrt{|(\langle\psi|\bar{A}\bar{B}^{\Psi}_n|\psi\rangle|^2} \geq |\langle\psi|(\bar{A}\bar{B})|\psi\rangle|\end{align}
%From the above we again get the following 
\begin{align} \Delta A\Delta B\geq \sum_n|\langle\Psi|\bar{A}\bar{B}^{\psi}_n|\Psi\rangle| \geq |\langle\Psi|\bar{A}\bar{B}|\Psi\rangle|.
\end{align}
This is the tighter uncertainty relation for pure quantum states \cite{Mondal2017}. Thus, we have proved that the tighter uncertainty relation for mixed quantum states reduces to that of the pure quantum states under the right conditions.
We can now optimise Eq.\eqref{tightuncrelation} over the set $\{\ket{\psi_n}\}$ to tighten the bound even further as follows
%Now again expanding this we obtain the optimized version of the tighter uncertainty relations for mixed states, the same way as in Schrodinger equation for mixed quantum states, we get the following.
\begin{align}
\Delta A\Delta B\geq \max_{\{|\psi_n\rangle\}}\left(\sum_n\sqrt{|\mathrm{Tr}(\bar{A}\rho\bar{A}\bar{B}^{\psi}_n\rho\bar{B}^{\psi}_n)|}\right),
\end{align}
where the expectation values are defined over the mixed state $\rho$.
%We note in the same way as stated above that the above inequality holds in any way we define $\bar{B}^{\Psi}_n$ as, i.e., either as $|\Psi_n\rangle\langle\Psi_n|\bar{B}$, or as $\bar{B}|\Psi_n\rangle\langle\Psi_n|$.
Thus, we have proved the mixed state version of the tighter uncertainty relation.
The essential method we have used here is the Cauchy-Schwarz inequality for two `real' vectors in one of the steps. Now we comment on how the same method can be used to derive the tighter quantum speed limit for mixed quantum states as well. We know that the derivation of the quantum speed limit for the mixed quantum states by Uhlmann uses the Cauchy-Schwarz inequality in deriving the main bound. We propose that if we use the Cauchy-Schwarz inequality for two real vectors in place of the usual Cauchy-Schwarz inequality there, we will get a tighter version of Uhlmann's quantum speed limit bound for mixed quantum states. However, we leave this direction for future research.}

\section{Tighter Quantum Speed Limit for pure quantum states} \label{TQSL for pure states}

\theorem{The time evolution of a quantum state $\ket{\Psi(t)}$ under a unitary operation generated by a Hamiltonian $H$ is bounded by the following inequality}
\begin{align}
\tau\geq\frac{\hbar s_0(\tau)}{2\Delta H}+\frac{2}{\Delta H}\int^\tau _0 \frac{K(t)}{ \sin s_0(t)}dt,
\end{align}
where we have the following quantities 
\begin{align}
& s_0(\tau) = 2\cos^{-1}|\langle\Psi(0)|\Psi(\tau)\rangle|,  \nonumber \\
& \Delta H^2 = \langle\Psi(t)|H^2|\Psi(t)\rangle -\langle\Psi(t)|H|\Psi(t)\rangle^2~\mathrm{and} \nonumber\\
& K(t) = \sum_n|\langle \Psi(t)|\bar{A}\bar{B}^{\psi}_n|\Psi(t)\rangle|-|\langle \Psi(t)|\bar{A}\bar{B}|\Psi(t)\rangle|\geq 0,\nonumber\\
&\mathrm{where}~~ A=|\Psi(0)\rangle\langle\Psi(0)|~~ \mathrm{and}~~B=H.
\end{align}
\proof {%Here, we derive a tighter quantum speed limit using the tighter uncertainty relation equation. 
Consider two non-commuting operators $A$ and $B$, the tighter uncertainty relation then gives
\begin{align}
%\Delta A^2\Delta B^2&\geq(\sum_n|\langle \Psi|\bar{A}\bar{B}^{\psi}_n|\Psi\rangle|)^2 \nonumber\\
\Delta A\Delta B&\geq \sum_n|\langle \Psi|\bar{A}\bar{B}^{\psi}_n|\Psi\rangle|, \label{tight}
\end{align}
where $\bar{A}=A-\langle A\rangle$ and $\bar{B}^{\psi}_n=|\psi_n\rangle\langle\psi_n|\bar{B}$, the average values $\langle A\rangle$ and $\langle B\rangle$ of the Hermitian operators $A$ and $B$, respectively, being defined with respect to the pure quantum state $|\Psi\rangle$. Also, we have 
\begin{align}
\sum_n|\langle \Psi|\bar{A}\bar{B}^{\psi}_n|\Psi\rangle|\geq |\langle \Psi|\bar{A}\bar{B}|\Psi\rangle|.
\end{align}
Now, we add and subtract $|\langle \Psi|\bar{A}\bar{B}|\Psi\rangle|$ to the RHS of Eq.\eqref{tight}
\begin{align}
\Delta A\Delta B\geq (\sum_n|\langle \Psi|\bar{A}\bar{B}^{\psi}_n|\Psi\rangle|-|\langle \Psi|\bar{A}\bar{B}|\Psi\rangle|) +|\langle \Psi|\bar{A}\bar{B}|\Psi\rangle|.\nonumber
\end{align}
We again note that the following equation holds
$$|\langle \Psi|\bar{A}\bar{B}|\Psi\rangle|^2= \frac{1}{4} |\langle \Psi|[A,B]|\Psi\rangle|^2 +|\frac{1}{2}\langle\{A,B\}\rangle-\langle A\rangle\langle B\rangle|^2.$$
Therefore, we have 
$$ |\langle \Psi|\bar{A}\bar{B}|\Psi\rangle|\geq \frac{1}{2} |\langle \Psi|[A,B]|\Psi\rangle|.$$
Using the above equation, we have the following uncertainty relation 
\begin{align}
\Delta A\Delta B \geq \frac{1}{2} |\langle \Psi|[A,B]|\Psi\rangle|+ K(t),\label{tightuncforpure}
\end{align}
where $K(t) = (\sum_n|\langle \Psi|\bar{A}\bar{B}^{\psi}_n|\Psi\rangle|-|\langle \Psi|\bar{A}\bar{B}|\Psi\rangle|) \geq 0$ which is time dependent via its dependence on the time evolved quantum state $|\Psi\rangle=|\Psi(t)\rangle=e^{-iHt}|\Psi(0)\rangle$. We will denote $K$ for $K(t)$ in short and will use this notation in the coming sections.
%Now, we take $A,B$ as the following 
%operators. (In the next paragraph, after the derivation of the tighter quantum speed limit, we will prove that $K\geq 0$.) 
Let us now consider the operators $A$ and $B$ as follows
\begin{align}
A=|\Psi(0)\rangle\langle\Psi(0)|~\mathrm{and}~~B=H.%|\Psi\rangle=|\Psi(t)\rangle
\end{align}
For the pure state projector $A=|\Psi(0)\rangle\langle\Psi(0)|$, we have $\langle A\rangle= |\langle\Psi(0)|\Psi(t)\rangle|^2 = \cos^2\frac{s_0(t)}{2}$, where $s_0(t)$ is called the Bargmann angle. 
%Now let $\sin^2\frac{s_0(t)}{2}=1-|\langle\Psi(0)|\Psi(t)\rangle|^2=1-\langle A\rangle$, where we easily note that for pure states $A=|\Psi(0)\rangle\langle\Psi(0)|$, $\langle A\rangle=\langle A^2\rangle=|\langle\Psi(0)|\Psi(t)\rangle|^2$.
Therefore, the variance of $A$ is given as
\begin{align}
\Delta A^2=\langle A^2\rangle-\langle A\rangle^2 =\frac{1}{4}\sin^{2} s_0(t).\label{deltaA}
\end{align}
%Also using the same equations, we also get the following relation
%\begin{align}
%\langle A\rangle=\langle\Psi(0)|\Psi(t)\rangle|^2=\cos^2\frac{s_0(t)}{2}
%\end{align}
The range of $s_0(t)$ is taken to be from $0$ to $\frac{\pi}{2}$. Using the equation of motion for the average of $A$, we have
\begin{align}
i\hbar \frac{d}{dt}\langle A\rangle=\langle\Psi(t)|[A,H]|\Psi(t)]\rangle.\end{align}
%Taking absolute values on both sides, we get
%\begin{align}|\hbar \frac{d}{dt}\langle A\rangle|=|\langle\Psi(t)|[A,H]|\Psi(t)]\rangle|.
%\end{align}
Now, using the expectation value of $A$ in terms of the Bargmann angle we have
\begin{align}
\left|\frac{d\langle A\rangle}{dt}\right| =\frac{1}{2}\sin s_0(t)\frac{ds_0}{dt}.\label{da/dt}
\end{align}
Therefore, putting the values of $\Delta A$ and $\Delta B$ explicitly in Eq.\eqref{tightuncforpure} and using Eq.\eqref{da/dt} we get  
\begin{align}
\frac{1}{2}\sin s_0(t) \Delta H \ge \frac{\hbar}{4}\sin s_0(t)\frac{ds_0}{dt}+K(t).
%\Rightarrow\Delta H=\frac{\hbar}{2}\frac{ds_0}{dt}+\frac{2K(t)}{\sin s_0(t)}
\end{align}
Now, integrating the above equation with respect to time we obtain the new tighter quantum speed limit bound as given by
\begin{align}
\label{tqsl1}
\tau\geq\frac {\hbar s_0(\tau)}{2\Delta H}+\frac{2}{\Delta H}\int^\tau _0 \frac{K(t)}{\sin s_0(t)}dt.
\end{align}
The first term on the RHS is the MT bound and the second term on the right is always positive, therefore the above equation gives a quantum speed limit always tighter than the MT bound. We also expect the above bound to perform better than the standard quantum speed limit since we have used the tighter uncertainty relation to derive the new quantum speed limit bound above. Note that the maximized or optimized speed limit is obtained by optimizing over the choice of complete basis vectors $\{|\psi_n\rangle\}$ for $n=1,...,d$ as follows 
\begin{align}
\label{tqsl2}
\tau\geq\max_{\{\ket{\psi_n}\}}\left[\frac {\hbar s_0(\tau)}{2\Delta H}+\frac{2}{\Delta H}\int^\tau _0 \frac{K(t)}{\sin s_0(t)}dt\right].
\end{align}}
Thus, Eq.~(\ref{tqsl1}) and  Eq.~(\ref{tqsl2}) constitute tighter quantum speed limits for arbitrary unitary evolutions of pure quantum states. The standard QSL such as the MT bound follows as a special case of the new bound.
%In the next paragraphs we compare our new bound with the MT bound for quantum speed limit.

\proposition{K(t) is positive semidefinite.}
\proof {The expression for $K(t)$ is given as follows:
\begin{align}
K(t)=(\sum_n|\langle \Psi|\bar{A}\bar{B}^{\psi}_n|\Psi\rangle|-|\langle \Psi|\bar{A}\bar{B}|\Psi\rangle|),
\end{align}
where we have $\bar{A}=A-\langle A\rangle$, $\bar{B}=B-\langle B\rangle$ and $\bar{B}^{\psi}_n=|\psi_n\rangle\langle\psi_n|\bar{B}$. Using the fact that the sum of the absolute values of complex numbers is greater than or equal to the absolute values of the sum of the complex numbers we have 
\begin{align}
\sum_n|\langle \Psi|\bar{A}\bar{B}^{\psi}_n|\Psi\rangle|&\geq |\sum_n\langle\Psi|\bar{A}\bar{B}^{\psi}_n|\Psi\rangle|\nonumber\\
 %Now we take the summation on the term on the RHS of the above equation inside and get the following equality
&=|\langle \Psi|\bar{A}\sum_n(\bar{B}^{\psi}_n)|\Psi\rangle| \nonumber\\
&=\langle\Psi|\bar{A}\sum_n(|\psi_n\rangle\langle\psi_n|\bar{B})|\Psi\rangle|\nonumber\\
&=\langle\Psi|\bar{A}\bar{B}|\Psi\rangle|,
\end{align}
where we have used the completeness relation $\sum_n|\psi_n\rangle\langle\psi_n|=\mathbb{I}$. Therefore, using the above inequality, we get $K(t)\geq 0$.} %The equality cases occur when we have the sum of the absolute values equal to the absolute values of the sum in the mathematical sense. 
%Let us now analyze the equality condition for $d=2$. Let $|\psi_1\rangle$ and $|\psi_2\rangle$ be the two basis vectors which are the eigenvectors of any Hermitian operator other than $A$ and $B$. Then for convenience we take 
%$\langle\Psi|\bar{A}\bar{B}^{\psi}_1|\Psi\rangle|=a $ and $\langle \Psi|\bar{A}\bar{B}^{\psi}_2|\Psi\rangle|=b $. Now we know that the triangle inequality $|a|+|b|\geq |a+b|$ becomes an equality if and only if $a=cb$ where $c$ is a positive real number. This depends on the initial quantum state and the Hamiltonian. The same also holds true for mixed states by similar arguements and algebra.

\section{Tighter quantum speed limit for mixed quantum states}  \label{TQSL for mixed states}
 
{\theorem{For a quantum state $\rho(t)$, the speed of unitary evolution generated by the Hamiltonian $H$ is bounded by the following inequality}}
\begin{widetext}
\begin{align}
&\tau\geq\Bigg[\frac{\hbar}{\Delta H}\left(\cos^{-1}(\sqrt{\mathrm{Tr}(\rho_0\rho_\tau)})-\cos^{-1}(\sqrt{\mathrm{Tr}(\rho_0^2)})\right)+\frac{1}{\sqrt{\mathrm{Tr}(\rho_0^2)}\Delta H}\int_0^{\tau}\frac{K(t)dt}{\cos \frac{s_0(t)}{2}\sqrt{1-\mathrm{Tr}(\rho_0^2)\cos^2 \frac{s_0(t)}{2}}}\Bigg],\nonumber\\ 
&\mathrm{and ~the~ optimized ~version~ is ~given~ as }\nonumber\\
&\tau\geq\max_{\{\ket{\psi_n}\}}\Bigg[\frac{\hbar}{\Delta H}\left(\cos^{-1}(\sqrt{\mathrm{Tr}(\rho_0\rho_\tau)})-\cos^{-1}(\sqrt{\mathrm{Tr}(\rho_0^2)})\right)+\frac{1}{\sqrt{\mathrm{Tr}(\rho_0^2)}\Delta H}\int_0^{\tau}\frac{K(t)dt}{\cos \frac{s_0(t)}{2}\sqrt{1-\mathrm{Tr}(\rho_0^2)\cos^2 \frac{s_0(t)}{2}}}\Bigg],\nonumber\\
&\mathrm{where ~we~have~the~following~definitions}\nonumber\\
&s_0(t)=2\cos^{-1}{\sqrt{\frac{\mathrm{Tr}(\rho_0\rho_t)}{\mathrm{Tr}(\rho_0^2)}}}, ~
\Delta H ^2=\mathrm{Tr}(\rho H^2)-(\mathrm{Tr}(\rho H))^2,~
K(t)=\sum_n\sqrt{|\mathrm{Tr}(\bar{\rho}_0\rho_t\bar{\rho}_0\bar{H}^{\psi}_n\rho_t\bar{H}^{\psi}_n)|} - |\mathrm{Tr}(\bar{\rho}_0\rho_t\bar{H})|.
\end{align}
\end{widetext}
{\proof{Consider two non-commuting operators $A$ and $B$, the tighter uncertainty relation for a mixed state $\rho$ using Eq.\eqref{tightuncrelation} is given by
\begin{align} 
\Delta A\Delta B\geq \sum_n\sqrt{|\mathrm{Tr}(\bar{A}\rho\bar{A}\bar{B}^{\psi}_n\rho\bar{B}^{\psi}_n)|}.
\end{align}
Now adding and subtracting $|\mathrm{Tr}(\bar{A}\rho\bar{B})|$ to the RHS of the above equation, we get
\begin{align}\nonumber 
\Delta A\Delta B\geq& \Big [\sum_n\sqrt{|\mathrm{Tr}(\bar{A}\rho\bar{A}\bar{B}^{\psi}_n\rho\bar{B}^{\psi}_n)|}- |\mathrm{Tr}(\bar{A}\rho\bar{B})|\Big ] \nonumber\\ 
&+|\mathrm{Tr}(\bar{A}\rho\bar{B})|. \label{tightuncspeed}
\end{align}
We will now analyze the term $|\mathrm{Tr}(\bar{A}\rho\bar{B})|$. For this we use a more convenient notation as $|\mathrm{Tr}(\bar{A}\rho\bar{B})|=|\langle \bar{A}\bar{B}\rangle|$. We note that the following equation holds for all mixed quantum states and where the expectation values denoted by the angled brackets are with respect to the mixed quantum state $\rho$, i.e.,
\begin{align}\nonumber
|\langle \bar{A}\bar{B}\rangle|^2= \frac{1}{4} |\langle[A,B]\rangle|^2 +|\frac{1}{2}\langle\{A,B\}\rangle-2\langle A\rangle\langle B\rangle|^2.
\end{align}
%The above relation can be proven as follows
%\begin{align}
%|\langle \bar{A}\bar{B}\rangle|^2= |\langle (A-\langle A\rangle)(B-\langle B\rangle)\rangle|^2\\ \nonumber
%= |\langle (AB-\langle A\rangle B-\langle B\rangle A+\langle B\rangle\langle A\rangle)\rangle|^2\\ \nonumber
%=|\langle (AB-\langle B\rangle\langle A\rangle)\rangle|^2\\ \nonumber
%=|\langle (\frac{1}{2}(AB-BA+AB+BA))-\langle B\rangle\langle A\rangle)\rangle|^2\\ \nonumber
%=|\langle (\frac{1}{2}([A,B]+\{A,B\}))-\langle B\rangle\langle A\rangle)\rangle|^2\\ 
%\end{align}
%Since $A,B$ are Hermitian operators, then $\langle[A,B]\rangle$ is imaginary and $\langle\{A,B\}\rangle$ is real and $\langle A\rangle\langle B\rangle$ is real. Therefore, from the above equation we get 
%\begin{align}
%|\langle \bar{A}\bar{B}\rangle|^2
%=\frac{1}{4}|\langle [A,B]\rangle|^2+\frac{1}{4}|\langle\{A,B\}\rangle))-2\langle B\rangle\langle A\rangle)\rangle|^2 
%\end{align}
Since both the terms on the R.H.S are positive, we have 
\begin{align}
%|\langle \bar{A}\bar{B}\rangle|^2
%\geq\frac{1}{4}|\langle [A,B]\rangle|^2\\ \nonumber 
|\langle \bar{A}\bar{B}\rangle|
\geq\frac{1}{2}|\langle [A,B]\rangle|.
\end{align}
Using Eq.\eqref{tightuncspeed} and the inequality from the above equation we get
\begin{align}
\Delta A\Delta B \geq \frac{1}{2} |\langle[A,B]\rangle| +K(t),
\end{align}
where $K(t)=\Big [\sum_n\sqrt{|\mathrm{Tr}(\bar{A}\rho\bar{A}\bar{B}^{\Psi}_n\rho\bar{B}^{\Psi}_n)|}- |\mathrm{Tr}(\bar{A}\rho\bar{B})|\Big ]$ is positive semidefinite using Eq.\eqref{tighterthanRobertSch}. Let us now take the operators $A$ and $B$ as follows $A=\rho(0)$ and $B=H$, and  $\rho\equiv\rho(t)=e^{-iHt}\rho(0)e^{iHt}$. 
The variance of the operator $A$ is then given by 
\begin{align}
\Delta A^2&=\mathrm{Tr}(\rho(0)^2\rho(t))-(\mathrm{Tr}(\rho(0)\rho(t)))^2\nonumber \\&=\mathrm{Tr}(\rho_0^2\rho_t)-(\mathrm{Tr}(\rho_0\rho_t))^2,
\end{align}
where we have used the notation $\rho(0)\equiv\rho_0$ and $\rho(t)\equiv\rho_t$. We can now take the following parametrization
\begin{align}
\langle A\rangle=\mathrm{Tr}(\rho_0\rho_t)=\mathrm{Tr}(\rho_0^2)\cos^2\frac{s_0(t)}{2}.\label{expecA}
\end{align}
Now, using the equation of motion for the average of $A$, we get 
\begin{align}\nonumber
\left|\hbar \frac{d}{dt}\langle A\rangle\right|=|\langle[A,H]\rangle|,
\end{align}
where the averages are all with respect to the mixed quantum state $\rho$ and $A$ has no explicit time dependence. Using Eq.\eqref{expecA} then, we get
\begin{align}
\left|\frac{d\langle A\rangle}{dt}\right| =\mathrm{Tr}(\rho_0^2)\frac{\sin s_0(t)}{2}\frac{ds_0}{dt}.
\end{align}
Therefore, putting the values of $A$ and $B$ explicitly in the above derived equations we get 
\begin{align}
\Delta A \Delta H\geq \mathrm{Tr}(\rho_0^2)\frac{\hbar\sin s_0(t)}{4}\frac{ds_0}{dt}+K(t).\label{productofdeltaAdeltaH}
\end{align}
Now let us analyse the structure of $\Delta A^2$ as follows
\begin{align}
\Delta A^2=\mathrm{Tr}(\rho_0^2\rho_t)-(\mathrm{Tr}(\rho_0\rho_t))^2.
\end{align}
Let $\{\ket{k}\}$ be the eigenbasis of $\rho_0$ then we have
\begin{align}
\rho_0=\sum_{k}\lambda_{k}|k\rangle\langle k| ~\mathrm{and}~
\rho_0^2=\sum_{k}\lambda_{k}^2|k\rangle\langle k|.
\end{align}
Using the above equation then we have the following quantities
\begin{align}
\mathrm{Tr}(\rho_0\rho_t)&=\sum_{k}\lambda_{k}\langle k|\rho_t|k\rangle ~~~\mathrm{and}\nonumber\\
\mathrm{Tr}(\rho_0^2\rho_t)&=\sum_{k}\lambda_{k}^2\langle k|\rho_t|k\rangle.
\end{align}
Since, we know that $0\leq\lambda_{k}^2\leq \lambda_{k}\leq 1~\forall ~k$ and also $\langle k|\rho_t|k\rangle\geq 0 ~\forall~k$ because $\rho_t$ is a positive operator. Therefore, we get 
\begin{align}
\mathrm{Tr}(\rho_0\rho_t)\geq\mathrm{Tr}(\rho_0^2\rho_t).
\end{align}
Adding $-(\mathrm{Tr}(\rho_0\rho_t))^2$ on both side of the above equation we get 
\begin{align}
\mathrm{Tr}(\rho_0\rho_t)-(\mathrm{Tr}(\rho_0\rho_t))^2&\geq 
\mathrm{Tr}(\rho_0^2\rho_t)-(\mathrm{Tr}(\rho_0\rho_t))^2\nonumber\\
&=\Delta A^2.
\end{align}
Now, using Eq.\eqref{expecA} we get
\begin{align}
\mathrm{Tr}(\rho_0^2)\cos^2 \frac{s_0(t)}{2}(1-\mathrm{Tr}(\rho_0^2)\cos^2 \frac{s_0(t)}{2})\geq\Delta A^2
\end{align}
Taking square root on both sides and multiplying by $\Delta H$ we get 
%\begin{align}
%\sqrt{\mathrm{Tr}(\rho_0^2)}\cos \frac{s_0(t)}{2}\sqrt{(1-\mathrm{Tr}(\rho_0^2)\cos^2 \frac{s_0(t)}{2})}\geq\Delta A
%\end{align}
%Therefore we have the following 
\begin{align}
\sqrt{\mathrm{Tr}(\rho_0^2)}\cos \frac{s_0(t)}{2}\sqrt{(1-\mathrm{Tr}(\rho_0^2)\cos^2 \frac{s_0(t)}{2})}\Delta H\geq \Delta A\Delta H.
\end{align}
%Putting this inequality in the tighter uncertainty relation for mixed quantum states we get the following 
%\begin{align}
%\sqrt{\mathrm{Tr}(\rho_0^2)}\cos \frac{s_0(t)}{2}\sqrt{(1-\mathrm{Tr}(\rho_0^2)\cos^2 \frac{s_0(t)}{2})}\Delta H\geq \\ \nonumber \Delta A\Delta H  \geq \mathrm{Tr}(\rho_0^2)\frac{\hbar\sin s_0(t)}{2}\frac{ds_0}{dt}+K(t)
%\end{align}
The above inequality using Eq.\eqref{productofdeltaAdeltaH} becomes
\begin{align}
\sqrt{\mathrm{Tr}(\rho_0^2)}\cos \frac{s_0(t)}{2}\sqrt{(1-\mathrm{Tr}(\rho_0^2)\cos^2 \frac{s_0(t)}{2})}\Delta H\nonumber\\ \geq \mathrm{Tr}(\rho_0^2)\frac{\hbar\sin s_0(t)}{4}\frac{ds_0}{dt}+K(t).
\end{align}
%From the above we get the following
%\begin{align}\nonumber 
 % dt \geq \frac{\hbar\sqrt{\mathrm{Tr}(\rho_0^2)}\sin s_0(t)}{2\cos \frac{s_0(t)}{2}\sqrt{(1-\mathrm{Tr}(\rho_0^2)\cos^2 \frac{s_0(t)}{2})}\Delta H}ds_0\\ \nonumber +\frac{K(t)}{\sqrt{\mathrm{Tr}(\rho_0^2)}\cos \frac{s_0(t)}{2}\sqrt{(1-\mathrm{Tr}(\rho_0^2)\cos^2 \frac{s_0(t)}{2})}\Delta H}dt
%\end{align}
Now, integrating the above equation with respect to t and $s_0(t)$ we obtain the new quantum speed limit bound for mixed quantum states as follows 
%\begin{align}\nonumber 
 % \int_0^{\tau}dt \geq \int_{s_0}^{s_\tau}\frac{\hbar\sqrt{\mathrm{Tr}(\rho_0^2)}\sin s_0(t)}{2\cos \frac{s_0(t)}{2}\sqrt{(1-\mathrm{Tr}(\rho_0^2)\cos^2 \frac{s_0(t)}{2})}\Delta H}ds_0\\ \nonumber +\int_0^{\tau}\frac{K(t)}{\sqrt{\mathrm{Tr}(\rho_0^2)}\cos \frac{s_0(t)}{2}\sqrt{(1-\mathrm{Tr}(\rho_0^2)\cos^2 \frac{s_0(t)}{2})}\Delta H}dt
%\end{align}
%Simplifying the above integration by regrouping the terms, we get 
\begin{align}\nonumber 
  \tau \geq \frac{\hbar\sqrt{\mathrm{Tr}(\rho_0^2)}}{4\Delta H}\int_{s_0(0)}^{s_0(\tau)}\frac{\sin s_0(t)}{\cos \frac{s_0(t)}{2}\sqrt{(1-\mathrm{Tr}(\rho_0^2)\cos^2 \frac{s_0(t)}{2})}}ds_0 \\ \nonumber +\frac{1}{\sqrt{\mathrm{Tr}(\rho_0^2)}\Delta H}\int_0^{\tau}\frac{K(t)}{\cos \frac{s_0(t)}{2}\sqrt{(1-\mathrm{Tr}(\rho_0^2)\cos^2 \frac{s_0(t)}{2})}}dt.
\end{align}
The first term on the right hand side can be integrated in the analytical form, so that we get the following relation
\begin{align}\nonumber 
\tau &\geq \frac{\hbar\sqrt{\mathrm{Tr}(\rho_0^2)}}{4\Delta H}\Big [-4\frac{\sin^{-1}(\sqrt{\mathrm{Tr}(\rho_0^2)}\cos\frac{s_0(t)}{2})}{\sqrt{\mathrm{Tr}(\rho_0^2)}}\Big]_{s_0(0)}^{s_0(\tau)} \\ \nonumber &+\frac{1}{\sqrt{\mathrm{Tr}(\rho_0^2)}\Delta H}\int_0^{\tau}\frac{K(t)}{\cos \frac{s_0(t)}{2}\sqrt{(1-\mathrm{Tr}(\rho_0^2)\cos^2 \frac{s_0(t)}{2})}}dt.
\end{align}
%Simplifying the above equation we get the following
%\begin{align}\nonumber 
%\tau &\geq \frac{\hbar}{\Delta H}\Big [\sin^{-1}(\sqrt{\mathrm{Tr}(\rho_0^2)}\cos\frac{s_0(0)}{2}) -\sin^{-1}(\sqrt{\mathrm{Tr}(\rho_0^2)}\cos\frac{s_0(\tau)}{2})\Big]\nonumber \\&+\frac{1}{\sqrt{\mathrm{Tr}(\rho_0^2)}\Delta H}\int_0^{\tau}\frac{K(t)}{\cos \frac{s_0(t)}{2}\sqrt{(1-\mathrm{Tr}(\rho_0^2)\cos^2 \frac{s_0(t)}{2})}}dt.\nonumber
%\end{align}
Now, we know that $\cos\frac{s_0(0)}{2}=1$ and $\sqrt{\mathrm{Tr}(\rho_0^2)}\cos\frac{s_0(\tau)}{2}=\sqrt{\mathrm{Tr}(\rho_0\rho_\tau)}$. Thus, putting these value in the above equation and simplifying, we get 
\begin{align}\nonumber 
\tau &\geq \frac{\hbar}{\Delta H}\Big [\sin^{-1}(\sqrt{\mathrm{Tr}(\rho_0^2)})-\sin^{-1}(\sqrt{\mathrm{Tr}(\rho_0\rho_\tau)})\Big] \\ \nonumber 
&+\frac{1}{\sqrt{\mathrm{Tr}(\rho_0^2)}\Delta H}\int_0^{\tau}\frac{K(t)}{\cos \frac{s_0(t)}{2}\sqrt{(1-\mathrm{Tr}(\rho_0^2)\cos^2 \frac{s_0(t)}{2})}}dt.
\end{align}
As before, $K(t)$ is always greater than or equal to zero in all cases. Other than that, the maximized version follows in the same way as in the case of mixed states without any further need of extra steps. Let us now check the bound for the case of pure quantum states as follows
\begin{align}
\rho_0=|\Psi(0)\rangle\langle\Psi(0)|, ~\rho_\tau=|\Psi(\tau)\rangle\langle\Psi(\tau)|.
\end{align}
In this case, our bound becomes the following 
\begin{align}\nonumber 
  \tau \geq \frac{\hbar}{\Delta H}\big(\sin^{-1}(1)-\sin^{-1}(\cos \frac{s_0(\tau)}{2})\big) +\frac{2}{\Delta H}\int_0^{\tau}\frac{K(t)dt}{\sin s_0(t)}.
\end{align}
Using $\sin^{-1}(1)=\frac{\pi}{2}$ and the following inverse trigonometric identity 
\begin{align}
\sin^{-1} x + \cos^{-1} x = \frac{\pi}{2},~ \forall~ x ~\in [-1,1],
\end{align}
we get the following
\begin{align}\nonumber 
  \tau \geq \Big [\hbar\big(\frac{\cos^{-1}(\cos \frac{s_0(\tau)}{2})}{\Delta H}\big)\Big] +\frac{2}{\Delta H}\int_0^{\tau}\frac{K(t)}{\sin s_0(t)}dt.
\end{align}
Therefore, this gives us the following quantum speed limit for pure quantum states
\begin{align}\nonumber 
  \tau \geq \frac{\hbar s_0(\tau)}{2\Delta H} +\frac{2}{\Delta H}\int_0^{\tau}\frac{K(t)}{\sin s_0(t)}dt.
\end{align}
Thus, the quantum speed limit bound for mixed quantum states reduces to that of the pure quantum states derived from the tighter uncertainty relation for pure quantum states in the appropriate limit. Note that in a more conventional notation using the trigonometric identity of inverses of $\sin$ and $\cos$ functions, we can write the bound as follows
\begin{align}\nonumber
\tau&\geq\Bigg[\frac{\hbar}{\Delta H}\left(\cos^{-1}(\sqrt{\mathrm{Tr}(\rho_0\rho_\tau)})-\cos^{-1}(\sqrt{\mathrm{Tr}(\rho_0^2)})\right)\\ \nonumber &+\frac{1}{\sqrt{\mathrm{Tr}(\rho_0^2)}\Delta H}\int_0^{\tau}\frac{K(t)dt}{\cos \frac{s_0(t)}{2}\sqrt{1-\mathrm{Tr}(\rho_0^2)\cos^2 \frac{s_0(t)}{2}}}\Bigg].
\end{align}
The optimized version can be expressed as
\begin{align}\nonumber
\tau&\geq\max_{\{\ket{\psi_n}\}}\Bigg[\frac{\hbar}{\Delta H}\left(\cos^{-1}(\sqrt{\mathrm{Tr}(\rho_0\rho_\tau)})-\cos^{-1}(\sqrt{\mathrm{Tr}(\rho_0^2)})\right)\\ \nonumber &+\frac{1}{\sqrt{\mathrm{Tr}(\rho_0^2)}\Delta H}\int_0^{\tau}\frac{K(t)dt}{\cos \frac{s_0(t)}{2}\sqrt{1-\mathrm{Tr}(\rho_0^2)\cos^2 \frac{s_0(t)}{2}}}\Bigg].
\end{align}
The quantum speed limit for mixed quantum states derived from the tighter uncertainty relation is another important result derived in the paper. The performance of this bound depends on the value of the second integration and it cannot be said a priori in a straightforward way in which cases it will perform better than the other existing bounds in the literature. As a result, we leave this direction for future research.
Next, we demonstrate the better performance of our bound over the MT bound and the bound in Eq.\eqref{better bound} with some examples in the case of pure quantum states.

\section{Examples} \label{examples}
In this section, we illustrate the tighter QSL for few examples where we see dramatic improvement over the standard QSL such as the MT bound.
In the first example, we discuss the tighter QSL for quantum sytsem whose dynamics is governed by random Hamiltonians. In the second and the third example, we discuss the tighter QSL for interacting systems of spins.

\subsection{Tighter QSL with random Hamiltonians}
In this subsection, we calculate and compare the tighter quantum speed limit bound with that of the MT bound using random Hamiltonians from the Gaussian Unitary ensemble (GUE). Random Hamiltonians drawn from one of the random matrix ensembles such as the GUE can be applicable to a large class of physically important models in quantum physics, quantum information and computation where long-range interactions are important. Recently, the random Hamiltonian setup as in the models of random quantum circuits have been used for analyzing features of quantum entanglement \cite{you2018}, universal properties of the out-of-time-ordered correlation function \cite{cotler2017,vijay2018,yoshida2019,gharibyan2017}, quantum entanglement tsunami \cite{nahum2017}, unitary design \cite{nicholas2019} and also measurement induced phase transitions \cite{skinner2019}. As a result, due to such physical applicability we study the performance of the tighter quantum speed limit bound for the random Hamiltonians.

Here, we state how we draw the random Hamiltonians and its mathematical properties. A random Hamiltonian is a $D\mathrm{x}D$ Hermitian operator $H$ drawn from a Gaussian Unitary Ensemble (GUE), described by the following probability distribution function 
$P(H)= Ce^{-\frac{D}{2}\mathrm{Tr}(H^2)}$, 
where $C$ is the normalization constant and the elements of $H$ are drawn from the Gaussian probability distribution. In this way $H$ is also Hermitian. A random Hamiltonian dynamics is a unitary time-evolution generated by a fixed (time-independent) GUE Hamiltonian. The matrices of this form can be drawn from mathematica in a straightforward way using routine procedure.
 
For the example considered here, we take the Hilbert space of dimension 3. The initial state is taken as $|\Psi(0) \rangle=  \sqrt{0.1}|0\rangle+\sqrt{0.2}|1\rangle+\sqrt{0.7}|2\rangle$. The random eigenbasis is taken as the set of eigenvectors of a random Hermitian operator obtained in the same way as the Hamiltonian, i.e., from the Gaussian Unitary Ensemble. In Figure \ref{fig:tqsl1}, we plot $\Delta=\tau_{tqsl}-\tau_{MT}$ vs the actual time $t$ of evolution, for 3 different random Hamiltonians. It is clear that $\Delta$ is always positive, showing that the tighter quantum speed limit bound always outperforms the MT bound. At $t=0$, all the values of $\Delta$ are zero because all the random Hamiltonians start with being identity at $t=0$. All the Hamiltonians taken here are time independent. As a result, we are able to show that with these Hamiltonians our bound outperforms the MT bound.
 \begin{figure}
\centering
\includegraphics[scale=0.625]{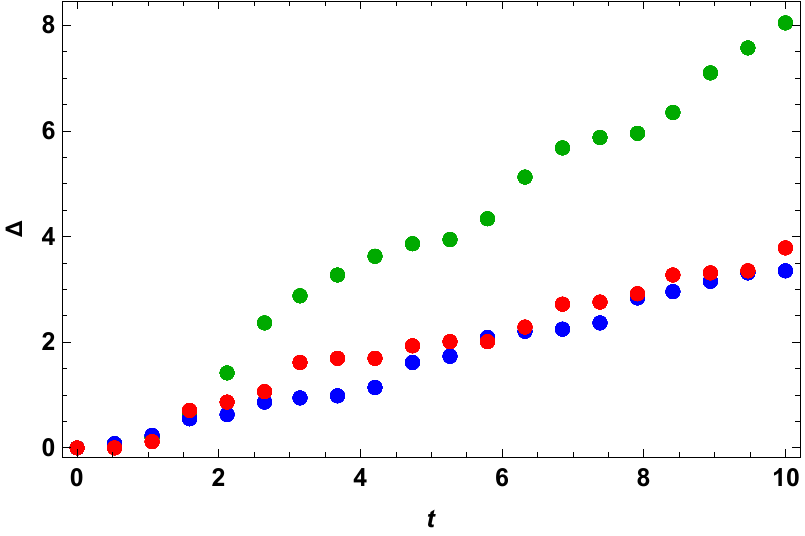}
\caption{The difference $\Delta$ between the tighter quantum speed limit $\tau_{tqsl}$ and the MT bound $\tau_{MT}$, obtained for 3 different random Hamiltonians obtained from a Gaussian Unitary Ensemble. As expected from the theory, the tighter quantum speed limit bound outperforms the MT bound always for these random Hamiltonians. The same holds for many other random Hamiltonian obtained in the same way from the Gaussian Unitary Ensemble. The initial state is taken as $\sqrt{0.1}|0\rangle+\sqrt{0.2}|1\rangle+\sqrt{0.7}|2\rangle$ in the three dimensional Hilbert space.}
\label{fig:tqsl1}
\end{figure}

\subsection{Interacting quantum systems of spins}

In this subsection, we work out another example to illustrate our bound. We choose the basis vectors to be from any Hermitian operator chosen from a GUE. We consider a chain of $M$ spins that evolve under the Hamiltonian
$H=\sum_{i=1}^M H_i +H_{int}$,
 \begin{figure}
\centering
\includegraphics[scale=0.675]{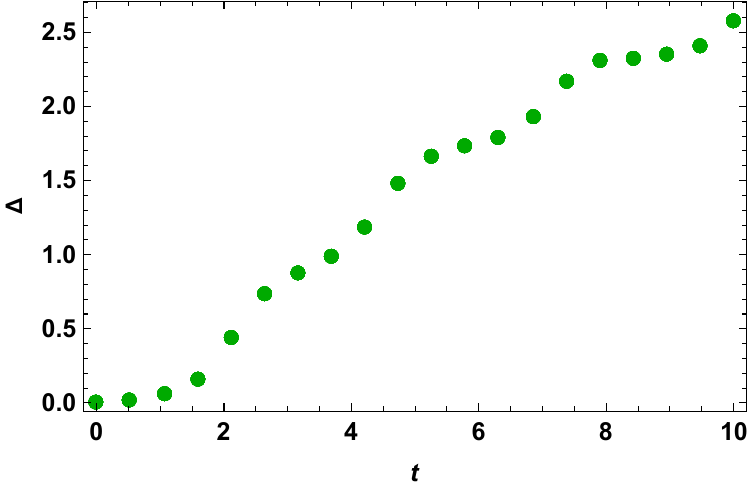}
\caption{The difference $\Delta$ between the tighter quantum speed limit $\tau_{tqsl}$ and the MT bound $\tau_{MT}$ obtained for the example in section VI (B).}
\label{fig:tqsl2}
\end{figure}
where we have $H_i=\hbar \omega_0(1-\sigma_x^i)$~\cite{Giovannetti2004}. Here, $H_i$ is the local Hamiltonian that evolves the individual spin systems independent of each other, whereas $H_{int}$ acts on subsystems jointly. Here, we assume that the interaction takes place in each of the $Q$ number of blocks present in the spin chain. Each block consists of $K$ number of spins. The $K$ spins in the $j^{th}$ block interact through the Hamiltonian $H_j$. We, therefore, have the following
\begin{align}
H_j=\hbar\omega(1-S_j),~~
H_{int}=\hbar\omega\sum_{j=1}^Q(1-S_j),\nonumber\\
\mathrm{where}~~
S_j =\sigma_x^{i_{1_j}}\otimes\sigma_x^{i_{2_j}} \otimes ....\otimes \sigma_x^{i_{k_j}}, j=1,2,..., Q.
\end{align}

Under this Hamiltonian, starting from a completely product state $|\Psi(0)\rangle$, the time evolved quantum state is of the following form 
\begin{align}\nonumber
|\Psi(t)\rangle=&C\bigotimes_{i=1}^M (\cos \omega_0t+i\sigma_x^i\sin\omega_0t)\\ \nonumber &\Pi_{j=1}^Q (\cos \omega t+iS_j\sin\omega t)|\Psi(0)\rangle,
\end{align}
where $C=e^{-i(\omega+\omega_0)t}$. For our case, we take two qubit system and the initial state as the product state $|0\rangle|0\rangle$, where $|0\rangle$ is the eigenstate of the operator $\sigma_z$. Also, we take the simplest case of a single block. We take this state as the initial state and evolve it under the Hamiltonian as stated above. The random eigenbasis is again taken as the set of eigenvectors of a random Hermitian operator obtained from the Gaussian Unitary Ensemble. In Figure \ref{fig:tqsl2}, we plot $\Delta=\tau_{tqsl}-\tau_{MT}$ vs the actual time $t$ of evolution. The figure clearly shows that our bound performs better than the MT bound.
\begin{figure}[h]
\centering
\includegraphics[scale=0.675]{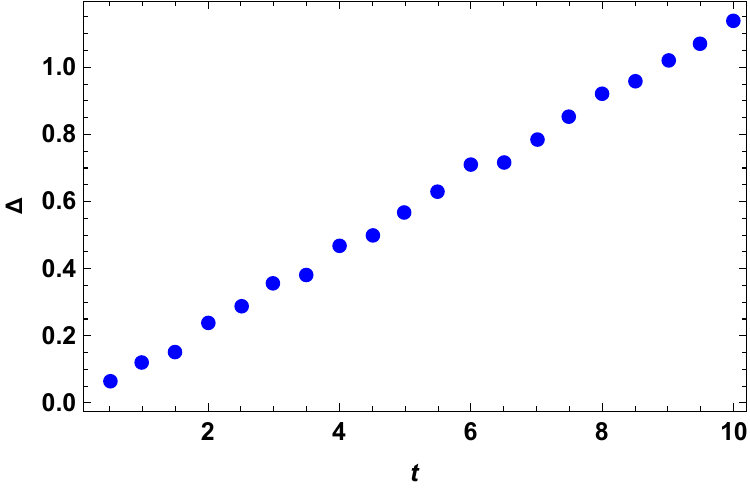}
\caption{The difference $\Delta$ between the tighter quantum speed limit $\tau_{tqsl}$ and the MT bound $\tau_{MT}$ obtained for the example in section VI (C).}
\label{fig:tqsl3}
\end{figure}

\subsection{Spin chains with nearest neighbour and next nearest neighbour interactions}

The Hamiltonian for Heisenberg model with nearest neighbor and next-nearest neighbor interaction can be written as follows 
\begin{eqnarray}
H=J_1\sum_i(\sigma_i^x\sigma_{i+1}^x+\sigma_i^y\sigma_{i+1}^y+\sigma_i^z\sigma_{i+1}^z) \nonumber \\
+J_2 \sum_i(\sigma_i^x\sigma_{i+2}^x+\sigma_i^y\sigma_{i+2}^y+\sigma_i^z\sigma_{i+2}^z). \label{heisenberghamiltonian}
\end{eqnarray}
This is a very well studied model of spin chains. Here, we work with the case of three qubits. As the initial state we take an entangled state in the seven dimensional subspace of the full Hilbert space as follows
\begin{align}|\Psi(0)\rangle=\begin{bmatrix}
\frac{1}{\sqrt{7}}&\frac{1}{\sqrt{7}}&\frac{1}{\sqrt{7}}&\frac{1}{\sqrt{7}}&\frac{1}{\sqrt{7}}&\frac{1}{\sqrt{7}}&0&\frac{1}{\sqrt{7}}\end{bmatrix}.
\end{align} 
This state is evolved by a time evolution operator generated by the Hamiltonian in Eq.\eqref{heisenberghamiltonian}.
The time evolved quantum state $\ket{\Psi(t)}$ is an entangled state in the full eight dimensional Hilbert space and is of the following form 
\begin{widetext}
{\small{\begin{align}\nonumber
\begin{bmatrix}
\frac{e^{-i(3 J_1 + J_2) t}}{\sqrt{7}}&
\frac{e^{-i(3 J_1 + J_2) t}}{\sqrt{7}}&
\frac{e^{-i(3 J_1 + J_2) t}}{\sqrt{7}}&
 \frac{(e^{-i (3 J_1 + J_2) t} (4 - e^{i6J_1t} (1 + 3 e^{i4J_2t})))}{(
  6 \sqrt{7})}&
\frac{e^{-i(3 J_1 + J_2) t}}{\sqrt{7}}~~~ ...~~~ &\\ 
~~~ ... ~~~\frac{(e^{-i (3 J_1 + J_2) t} (2 + e^{i6J_1t}))}{(3 \sqrt{7})}&
 \frac{(e^{-i (3 J_1 + J_2) t} (4 + 
     e^{i6J_1t} (-1 + 3 e^{i4J_2t})))}{(6 \sqrt{7})}&
\frac{e^{-i(3 J_1 + J_2) t}}{\sqrt{7}}
\end{bmatrix}.
\end{align}}}
\end{widetext}
The random eigenbasis is again taken as the set of eigenvectors of a random Hermitian operator obtained from the Gaussian Unitary Ensemble. In Figure \ref{fig:tqsl3}, we plot $\Delta=\tau_{tqsl}-\tau_{MT}$ vs the actual time $t$ of evolution. The figure clearly shows that our bound performs better than the MT bound.

\subsubsection{Comparison with the bound in \texorpdfstring{Eq.\eqref{better bound}}{}  using random quantum states as the initial states.}

In this subsection, we compare the quantum speed limit bound for our case with that of the Eq.\eqref{better bound}, for the case of random initial pure states obtained from the Gaussian random numbers and normalizing the obtained vector. We obtain the difference of our bound with respect to the other bound for four different time slots all using the ten different random initial quantum states for the same Hamiltonian as in the above section, i.e., Heisenberg spin chain with nearest neighbour and next nearest neighbour interaction for the case of three qubits. This shows that in all these diverse cases, our bound performs much better than most of the earlier bounds proposed in many different conditions. We expect that our bound will also perform better than most of the earlier bounds for the case of mixed quantum states. 

\begin{figure}[h]
\centering
\includegraphics[scale=0.675]{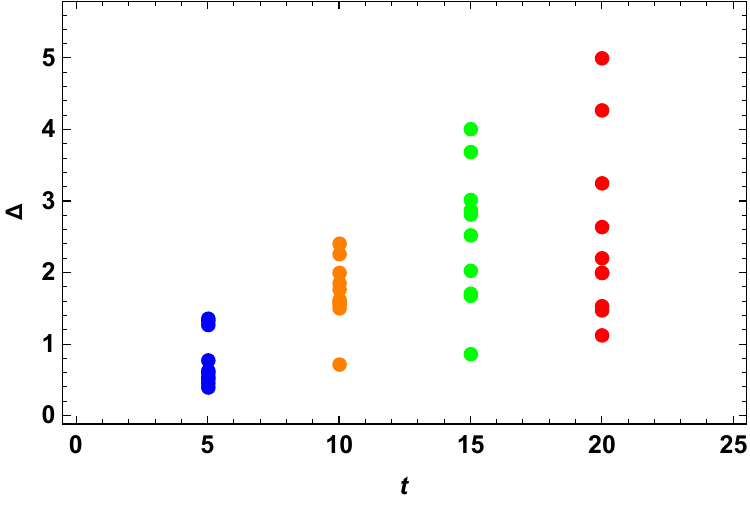}
\caption{The difference $\Delta$ between the tighter quantum speed limit $\tau_{tqsl}$ and the second bound $\tau_{2}$ obtained for the example in section VI (C(1)) for Hamitonian evolution of pure quantum states.}
\label{fig:tqsl4}
\end{figure}

\section{Conclusions and Future Directions}  \label{conclusion}

In this work, we have derived a tighter quantum speed limit. We first derived the mixed state generalization of the tighter uncertainty relation for pure quantum states. Using the tighter uncertainty relations for the pure and mixed quantum states, we have derived the tighter quantum speed limit bounds for pure and mixed quantum states, respectively. We have shown that the new bound performs better than the MT bound. Also, the tighter quantum speed limit bound has been shown to coincide with that of the pure quantum states in the appropriate limit. Hereafter, we have shown numerically using random Hamiltonians obtained from Gaussian Unitary Ensemble that our bound performs better than the MT speed limit bound.
Also, we have shown the better performance of our bound in some analytical examples involving spin chain and spin chain interactions. Apart from these, we have also shown the better performance of our bound than another bound in the current literature using random quantum states as initial quantum states undergoing Hamiltonian evolution involving spin chains. Since, we have shown that our bound is always better than the MT bound in all cases, therefore, all the cases where the MT bound performs better than the Margolus-Levitin bound, our new bound also performs better than the Margolus- Levitin bound in those cases. It remains a subject of future investigation to compare our bound with the Margolus-Levitin bound where MT bound is weaker than the Margolus-Levitin bound. In future, it will be interesting to see how much the speed limit bounds can be improved by optimization over different choice of sets of orthonormal bases and which basis set will be the optimal one in deriving the optimal quantum speed limit bound. Also, one could generalize our tighter quantum speed limit bound to the case of mixed quantum states for open systems dynamics.

We believe that the tighter quantum speed limit derived here will have important applications in quantum computation, quantum information and quantum control.

\section*{ACKNOWLEDGEMENTS}
S. B. acknowledges support from the National Research Foundation of Korea (2020M3E4A1079939, 2022M3K4A1094774) and the KIST institutional program (2E31531). A.S. acknowledges the support from the INFOSYS scholarship. A. K. P. acknowledges the support from the QUEST Grant Q-117 and J C Bose grant from the Department of Science and Technology, India.

\bibliography{ref}

%apsrev4-2.bst 2019-01-14 (MD) hand-edited version of apsrev4-1.bst
%Control: key (0)
%Control: author (8) initials jnrlst
%Control: editor formatted (1) identically to author
%Control: production of article title (0) allowed
%Control: page (0) single
%Control: year (1) truncated
%Control: production of eprint (0) enabled
\begin{thebibliography}{83}%
\makeatletter
\providecommand \@ifxundefined [1]{%
 \@ifx{#1\undefined}
}%
\providecommand \@ifnum [1]{%
 \ifnum #1\expandafter \@firstoftwo
 \else \expandafter \@secondoftwo
 \fi
}%
\providecommand \@ifx [1]{%
 \ifx #1\expandafter \@firstoftwo
 \else \expandafter \@secondoftwo
 \fi
}%
\providecommand \natexlab [1]{#1}%
\providecommand \enquote  [1]{``#1''}%
\providecommand \bibnamefont  [1]{#1}%
\providecommand \bibfnamefont [1]{#1}%
\providecommand \citenamefont [1]{#1}%
\providecommand \href@noop [0]{\@secondoftwo}%
\providecommand \href [0]{\begingroup \@sanitize@url \@href}%
\providecommand \@href[1]{\@@startlink{#1}\@@href}%
\providecommand \@@href[1]{\endgroup#1\@@endlink}%
\providecommand \@sanitize@url [0]{\catcode `\\12\catcode `\$12\catcode
  `\&12\catcode `\#12\catcode `\^12\catcode `\_12\catcode `\%12\relax}%
\providecommand \@@startlink[1]{}%
\providecommand \@@endlink[0]{}%
\providecommand \url  [0]{\begingroup\@sanitize@url \@url }%
\providecommand \@url [1]{\endgroup\@href {#1}{\urlprefix }}%
\providecommand \urlprefix  [0]{URL }%
\providecommand \Eprint [0]{\href }%
\providecommand \doibase [0]{https://doi.org/}%
\providecommand \selectlanguage [0]{\@gobble}%
\providecommand \bibinfo  [0]{\@secondoftwo}%
\providecommand \bibfield  [0]{\@secondoftwo}%
\providecommand \translation [1]{[#1]}%
\providecommand \BibitemOpen [0]{}%
\providecommand \bibitemStop [0]{}%
\providecommand \bibitemNoStop [0]{.\EOS\space}%
\providecommand \EOS [0]{\spacefactor3000\relax}%
\providecommand \BibitemShut  [1]{\csname bibitem#1\endcsname}%
\let\auto@bib@innerbib\@empty
%</preamble>
\bibitem [{\citenamefont {Heisenberg}(1927)}]{Heisenberg1927}%
  \BibitemOpen
  \bibfield  {author} {\bibinfo {author} {\bibfnamefont {W.}~\bibnamefont
  {Heisenberg}},\ }\bibfield  {title} {\bibinfo {title} {Über den
  anschaulichen inhalt der quantentheoretischen kinematik und mechanik},\
  }\href {https://doi.org/https://doi.org/10.1007/BF01397280} {\bibfield
  {journal} {\bibinfo  {journal} {Zeitschrift für Physik}\ }\textbf {\bibinfo
  {volume} {43}},\ \bibinfo {pages} {172} (\bibinfo {year} {1927})}\BibitemShut
  {NoStop}%
\bibitem [{\citenamefont {Robertson}(1929)}]{Robertson1929}%
  \BibitemOpen
  \bibfield  {author} {\bibinfo {author} {\bibfnamefont {H.~P.}\ \bibnamefont
  {Robertson}},\ }\bibfield  {title} {\bibinfo {title} {The uncertainty
  principle},\ }\href {https://doi.org/10.1103/PhysRev.34.163} {\bibfield
  {journal} {\bibinfo  {journal} {Physical Review}\ }\textbf {\bibinfo {volume}
  {34}},\ \bibinfo {pages} {163} (\bibinfo {year} {1929})}\BibitemShut
  {NoStop}%
\bibitem [{\citenamefont {Aharonov}\ and\ \citenamefont
  {Bohm}(1961)}]{Aharonov1961}%
  \BibitemOpen
  \bibfield  {author} {\bibinfo {author} {\bibfnamefont {Y.}~\bibnamefont
  {Aharonov}}\ and\ \bibinfo {author} {\bibfnamefont {D.}~\bibnamefont
  {Bohm}},\ }\bibfield  {title} {\bibinfo {title} {Time in the quantum theory
  and the uncertainty relation for time and energy},\ }\href
  {https://doi.org/10.1103/PhysRev.122.1649} {\bibfield  {journal} {\bibinfo
  {journal} {Physical Review}\ }\textbf {\bibinfo {volume} {122}},\ \bibinfo
  {pages} {1649} (\bibinfo {year} {1961})}\BibitemShut {NoStop}%
\bibitem [{\citenamefont {Aharonov}\ \emph {et~al.}(2002)\citenamefont
  {Aharonov}, \citenamefont {Massar},\ and\ \citenamefont
  {Popescu}}]{Aharonov2002}%
  \BibitemOpen
  \bibfield  {author} {\bibinfo {author} {\bibfnamefont {Y.}~\bibnamefont
  {Aharonov}}, \bibinfo {author} {\bibfnamefont {S.}~\bibnamefont {Massar}},\
  and\ \bibinfo {author} {\bibfnamefont {S.}~\bibnamefont {Popescu}},\
  }\bibfield  {title} {\bibinfo {title} {Measuring energy, estimating
  hamiltonians, and the time-energy uncertainty relation},\ }\href
  {https://doi.org/10.1103/PhysRevA.66.052107} {\bibfield  {journal} {\bibinfo
  {journal} {Physical Review A}\ }\textbf {\bibinfo {volume} {66}},\ \bibinfo
  {pages} {052107} (\bibinfo {year} {2002})}\BibitemShut {NoStop}%
\bibitem [{\citenamefont {Busch}(2008)}]{Busch2008}%
  \BibitemOpen
  \bibfield  {author} {\bibinfo {author} {\bibfnamefont {P.}~\bibnamefont
  {Busch}},\ }\bibinfo {title} {The time--energy uncertainty relation},\ in\
  \href {https://doi.org/10.1007/978-3-540-73473-4_3} {\emph {\bibinfo
  {booktitle} {Time in Quantum Mechanics}}},\ \bibinfo {editor} {edited by\
  \bibinfo {editor} {\bibfnamefont {J.}~\bibnamefont {Muga}}, \bibinfo {editor}
  {\bibfnamefont {R.~S.}\ \bibnamefont {Mayato}},\ and\ \bibinfo {editor}
  {\bibfnamefont {{\'I}.}~\bibnamefont {Egusquiza}}}\ (\bibinfo  {publisher}
  {Springer Berlin Heidelberg},\ \bibinfo {address} {Berlin, Heidelberg},\
  \bibinfo {year} {2008})\ pp.\ \bibinfo {pages} {73--105}\BibitemShut
  {NoStop}%
\bibitem [{\citenamefont {Maccone}\ and\ \citenamefont
  {Pati}(2014)}]{Pati2014}%
  \BibitemOpen
  \bibfield  {author} {\bibinfo {author} {\bibfnamefont {L.}~\bibnamefont
  {Maccone}}\ and\ \bibinfo {author} {\bibfnamefont {A.~K.}\ \bibnamefont
  {Pati}},\ }\bibfield  {title} {\bibinfo {title} {Stronger uncertainty
  relations for all incompatible observables},\ }\href
  {https://doi.org/10.1103/PhysRevLett.113.260401} {\bibfield  {journal}
  {\bibinfo  {journal} {Physical Review Letters}\ }\textbf {\bibinfo {volume}
  {113}},\ \bibinfo {pages} {260401} (\bibinfo {year} {2014})}\BibitemShut
  {NoStop}%
\bibitem [{\citenamefont {Mondal}\ \emph {et~al.}(2017)\citenamefont {Mondal},
  \citenamefont {Bagchi},\ and\ \citenamefont {Pati}}]{Mondal2017}%
  \BibitemOpen
  \bibfield  {author} {\bibinfo {author} {\bibfnamefont {D.}~\bibnamefont
  {Mondal}}, \bibinfo {author} {\bibfnamefont {S.}~\bibnamefont {Bagchi}},\
  and\ \bibinfo {author} {\bibfnamefont {A.~K.}\ \bibnamefont {Pati}},\
  }\bibfield  {title} {\bibinfo {title} {Tighter uncertainty and reverse
  uncertainty relations},\ }\href {https://doi.org/10.1103/PhysRevA.95.052117}
  {\bibfield  {journal} {\bibinfo  {journal} {Physical Review A}\ }\textbf
  {\bibinfo {volume} {95}},\ \bibinfo {pages} {052117} (\bibinfo {year}
  {2017})}\BibitemShut {NoStop}%
\bibitem [{\citenamefont {Xiao}\ \emph {et~al.}(2020)\citenamefont {Xiao},
  \citenamefont {Fan}, \citenamefont {Wang}, \citenamefont {Pati},\ and\
  \citenamefont {Xue}}]{xiao2020}%
  \BibitemOpen
  \bibfield  {author} {\bibinfo {author} {\bibfnamefont {L.}~\bibnamefont
  {Xiao}}, \bibinfo {author} {\bibfnamefont {B.}~\bibnamefont {Fan}}, \bibinfo
  {author} {\bibfnamefont {K.}~\bibnamefont {Wang}}, \bibinfo {author}
  {\bibfnamefont {A.~K.}\ \bibnamefont {Pati}},\ and\ \bibinfo {author}
  {\bibfnamefont {P.}~\bibnamefont {Xue}},\ }\bibfield  {title} {\bibinfo
  {title} {Direct experimental test of forward and reverse uncertainty
  relations},\ }\href {https://doi.org/10.1103/PhysRevResearch.2.023106}
  {\bibfield  {journal} {\bibinfo  {journal} {Physical Review Research}\
  }\textbf {\bibinfo {volume} {2}},\ \bibinfo {pages} {023106} (\bibinfo {year}
  {2020})}\BibitemShut {NoStop}%
\bibitem [{\citenamefont {Mandelstam}\ and\ \citenamefont
  {Tamm}(1945)}]{Mandelstam1945}%
  \BibitemOpen
  \bibfield  {author} {\bibinfo {author} {\bibfnamefont {L.}~\bibnamefont
  {Mandelstam}}\ and\ \bibinfo {author} {\bibfnamefont {I.}~\bibnamefont
  {Tamm}},\ }\bibfield  {title} {\bibinfo {title} {The {U}ncertainty {R}elation
  {B}etween {E}nergy and {T}ime in {N}on-relativistic {Q}uantum {M}echanics},\
  }\href {https://doi.org/10.1007/978-3-642-74626-0_8} {\bibfield  {journal}
  {\bibinfo  {journal} {J. Phys. (USSR)}\ }\textbf {\bibinfo {volume} {9}},\
  \bibinfo {pages} {249} (\bibinfo {year} {1945})}\BibitemShut {NoStop}%
\bibitem [{\citenamefont {Margolus}\ and\ \citenamefont
  {Levitin}(1998)}]{Margolus1998}%
  \BibitemOpen
  \bibfield  {author} {\bibinfo {author} {\bibfnamefont {N.}~\bibnamefont
  {Margolus}}\ and\ \bibinfo {author} {\bibfnamefont {L.~B.}\ \bibnamefont
  {Levitin}},\ }\bibfield  {title} {\bibinfo {title} {The maximum speed of
  dynamical evolution},\ }\href
  {https://doi.org/https://doi.org/10.1016/S0167-2789(98)00054-2} {\bibfield
  {journal} {\bibinfo  {journal} {Physica D: Nonlinear Phenomena}\ }\textbf
  {\bibinfo {volume} {120}},\ \bibinfo {pages} {188} (\bibinfo {year}
  {1998})}\BibitemShut {NoStop}%
\bibitem [{\citenamefont {Anandan}\ and\ \citenamefont
  {Aharonov}(1990)}]{Anandan1990}%
  \BibitemOpen
  \bibfield  {author} {\bibinfo {author} {\bibfnamefont {J.}~\bibnamefont
  {Anandan}}\ and\ \bibinfo {author} {\bibfnamefont {Y.}~\bibnamefont
  {Aharonov}},\ }\bibfield  {title} {\bibinfo {title} {Geometry of quantum
  evolution},\ }\href {https://doi.org/10.1103/PhysRevLett.65.1697} {\bibfield
  {journal} {\bibinfo  {journal} {Physical Review Letters}\ }\textbf {\bibinfo
  {volume} {65}},\ \bibinfo {pages} {1697} (\bibinfo {year}
  {1990})}\BibitemShut {NoStop}%
\bibitem [{\citenamefont {Levitin}\ and\ \citenamefont
  {Toffoli}(2009)}]{Levitin2009}%
  \BibitemOpen
  \bibfield  {author} {\bibinfo {author} {\bibfnamefont {L.~B.}\ \bibnamefont
  {Levitin}}\ and\ \bibinfo {author} {\bibfnamefont {T.}~\bibnamefont
  {Toffoli}},\ }\bibfield  {title} {\bibinfo {title} {Fundamental {L}imit on
  the {R}ate of {Q}uantum {D}ynamics: The {U}nified {B}ound {I}s {T}ight},\
  }\href {https://doi.org/10.1103/PhysRevLett.103.160502} {\bibfield  {journal}
  {\bibinfo  {journal} {Physical Review Letters}\ }\textbf {\bibinfo {volume}
  {103}},\ \bibinfo {pages} {160502} (\bibinfo {year} {2009})}\BibitemShut
  {NoStop}%
\bibitem [{\citenamefont {Gislason}\ \emph {et~al.}(1985)\citenamefont
  {Gislason}, \citenamefont {Sabelli},\ and\ \citenamefont
  {Wood}}]{Gislason1956}%
  \BibitemOpen
  \bibfield  {author} {\bibinfo {author} {\bibfnamefont {E.~A.}\ \bibnamefont
  {Gislason}}, \bibinfo {author} {\bibfnamefont {N.~H.}\ \bibnamefont
  {Sabelli}},\ and\ \bibinfo {author} {\bibfnamefont {J.~W.}\ \bibnamefont
  {Wood}},\ }\bibfield  {title} {\bibinfo {title} {New form of the time-energy
  uncertainty relation},\ }\href {https://doi.org/10.1103/PhysRevA.31.2078}
  {\bibfield  {journal} {\bibinfo  {journal} {Physical Review A}\ }\textbf
  {\bibinfo {volume} {31}},\ \bibinfo {pages} {2078} (\bibinfo {year}
  {1985})}\BibitemShut {NoStop}%
\bibitem [{\citenamefont {Eberly}\ and\ \citenamefont
  {Singh}(1973)}]{Eberly1973}%
  \BibitemOpen
  \bibfield  {author} {\bibinfo {author} {\bibfnamefont {J.~H.}\ \bibnamefont
  {Eberly}}\ and\ \bibinfo {author} {\bibfnamefont {L.~P.~S.}\ \bibnamefont
  {Singh}},\ }\bibfield  {title} {\bibinfo {title} {Time {O}perators, {P}artial
  {S}tationarity, and the {E}nergy-{T}ime {U}ncertainty {R}elation},\ }\href
  {https://doi.org/10.1103/PhysRevD.7.359} {\bibfield  {journal} {\bibinfo
  {journal} {Physical Review D}\ }\textbf {\bibinfo {volume} {7}},\ \bibinfo
  {pages} {359} (\bibinfo {year} {1973})}\BibitemShut {NoStop}%
\bibitem [{\citenamefont {Bauer}\ and\ \citenamefont
  {Mello}(1978)}]{Bauer1978}%
  \BibitemOpen
  \bibfield  {author} {\bibinfo {author} {\bibfnamefont {M.}~\bibnamefont
  {Bauer}}\ and\ \bibinfo {author} {\bibfnamefont {P.}~\bibnamefont {Mello}},\
  }\bibfield  {title} {\bibinfo {title} {The time-energy uncertainty
  relation},\ }\href
  {https://doi.org/https://doi.org/10.1016/0003-4916(78)90223-3} {\bibfield
  {journal} {\bibinfo  {journal} {Annals of Physics}\ }\textbf {\bibinfo
  {volume} {111}},\ \bibinfo {pages} {38} (\bibinfo {year} {1978})}\BibitemShut
  {NoStop}%
\bibitem [{\citenamefont {Bhattacharyya}(1983)}]{Bhattacharyya1983}%
  \BibitemOpen
  \bibfield  {author} {\bibinfo {author} {\bibfnamefont {K.}~\bibnamefont
  {Bhattacharyya}},\ }\bibfield  {title} {\bibinfo {title} {Quantum decay and
  the {M}andelstam-{T}amm-energy inequality},\ }\href
  {https://doi.org/10.1088/0305-4470/16/13/021} {\bibfield  {journal} {\bibinfo
   {journal} {Journal of Physics A: Mathematical and General}\ }\textbf
  {\bibinfo {volume} {16}},\ \bibinfo {pages} {2993} (\bibinfo {year}
  {1983})}\BibitemShut {NoStop}%
\bibitem [{\citenamefont {Leubner}\ and\ \citenamefont
  {Kiener}(1985)}]{Leubner1985}%
  \BibitemOpen
  \bibfield  {author} {\bibinfo {author} {\bibfnamefont {C.}~\bibnamefont
  {Leubner}}\ and\ \bibinfo {author} {\bibfnamefont {C.}~\bibnamefont
  {Kiener}},\ }\bibfield  {title} {\bibinfo {title} {Improvement of the
  {E}berly-{S}ingh time-energy inequality by combination with the
  {M}andelstam-{T}amm approach},\ }\href
  {https://doi.org/10.1103/PhysRevA.31.483} {\bibfield  {journal} {\bibinfo
  {journal} {Physical Review A}\ }\textbf {\bibinfo {volume} {31}},\ \bibinfo
  {pages} {483} (\bibinfo {year} {1985})}\BibitemShut {NoStop}%
\bibitem [{\citenamefont {Vaidman}(1992)}]{Vaidman1992}%
  \BibitemOpen
  \bibfield  {author} {\bibinfo {author} {\bibfnamefont {L.}~\bibnamefont
  {Vaidman}},\ }\bibfield  {title} {\bibinfo {title} {Minimum time for the
  evolution to an orthogonal quantum state},\ }\href
  {https://doi.org/10.1119/1.16940} {\bibfield  {journal} {\bibinfo  {journal}
  {American journal of physics}\ }\textbf {\bibinfo {volume} {60}},\ \bibinfo
  {pages} {182} (\bibinfo {year} {1992})}\BibitemShut {NoStop}%
\bibitem [{\citenamefont {Uhlmann}(1992)}]{Uhlmann1992}%
  \BibitemOpen
  \bibfield  {author} {\bibinfo {author} {\bibfnamefont {A.}~\bibnamefont
  {Uhlmann}},\ }\bibfield  {title} {\bibinfo {title} {An energy dispersion
  estimate},\ }\href
  {https://doi.org/https://doi.org/10.1016/0375-9601(92)90555-Z} {\bibfield
  {journal} {\bibinfo  {journal} {Physics Letters A}\ }\textbf {\bibinfo
  {volume} {161}},\ \bibinfo {pages} {329} (\bibinfo {year}
  {1992})}\BibitemShut {NoStop}%
\bibitem [{\citenamefont {Uffink}(1993)}]{Uffink1993}%
  \BibitemOpen
  \bibfield  {author} {\bibinfo {author} {\bibfnamefont {J.~B.}\ \bibnamefont
  {Uffink}},\ }\bibfield  {title} {\bibinfo {title} {The rate of evolution of a
  quantum state},\ }\href {https://doi.org/10.1119/1.17368} {\bibfield
  {journal} {\bibinfo  {journal} {American Journal of Physics}\ }\textbf
  {\bibinfo {volume} {61}},\ \bibinfo {pages} {935} (\bibinfo {year}
  {1993})}\BibitemShut {NoStop}%
\bibitem [{\citenamefont {Pfeifer}\ and\ \citenamefont
  {Fr\"ohlich}(1995)}]{Pfeifer1995}%
  \BibitemOpen
  \bibfield  {author} {\bibinfo {author} {\bibfnamefont {P.}~\bibnamefont
  {Pfeifer}}\ and\ \bibinfo {author} {\bibfnamefont {J.}~\bibnamefont
  {Fr\"ohlich}},\ }\bibfield  {title} {\bibinfo {title} {Generalized
  time-energy uncertainty relations and bounds on lifetimes of resonances},\
  }\href {https://doi.org/10.1103/RevModPhys.67.759} {\bibfield  {journal}
  {\bibinfo  {journal} {Reviews of Modern Physics}\ }\textbf {\bibinfo {volume}
  {67}},\ \bibinfo {pages} {759} (\bibinfo {year} {1995})}\BibitemShut
  {NoStop}%
\bibitem [{\citenamefont {Horesh}\ and\ \citenamefont
  {Mann}(1998)}]{Horesh1998}%
  \BibitemOpen
  \bibfield  {author} {\bibinfo {author} {\bibfnamefont {N.}~\bibnamefont
  {Horesh}}\ and\ \bibinfo {author} {\bibfnamefont {A.}~\bibnamefont {Mann}},\
  }\bibfield  {title} {\bibinfo {title} {Intelligent states for the {A}nandan -
  {A}haronov parameter-based uncertainty relation},\ }\href
  {https://doi.org/10.1088/0305-4470/31/36/003} {\bibfield  {journal} {\bibinfo
   {journal} {Journal of Physics A: Mathematical and General}\ }\textbf
  {\bibinfo {volume} {31}},\ \bibinfo {pages} {L609} (\bibinfo {year}
  {1998})}\BibitemShut {NoStop}%
\bibitem [{\citenamefont {Pati}(1999)}]{AKPati1999}%
  \BibitemOpen
  \bibfield  {author} {\bibinfo {author} {\bibfnamefont {A.~K.}\ \bibnamefont
  {Pati}},\ }\bibfield  {title} {\bibinfo {title} {Uncertainty relation of
  {A}nandan–{A}haronov and intelligent states},\ }\href
  {https://doi.org/https://doi.org/10.1016/S0375-9601(99)00701-X} {\bibfield
  {journal} {\bibinfo  {journal} {Physics Letters A}\ }\textbf {\bibinfo
  {volume} {262}},\ \bibinfo {pages} {296} (\bibinfo {year}
  {1999})}\BibitemShut {NoStop}%
\bibitem [{\citenamefont {S\"oderholm}\ \emph {et~al.}(1999)\citenamefont
  {S\"oderholm}, \citenamefont {Bj\"ork}, \citenamefont {Tsegaye},\ and\
  \citenamefont {Trifonov}}]{Soderholm1999}%
  \BibitemOpen
  \bibfield  {author} {\bibinfo {author} {\bibfnamefont {J.}~\bibnamefont
  {S\"oderholm}}, \bibinfo {author} {\bibfnamefont {G.}~\bibnamefont
  {Bj\"ork}}, \bibinfo {author} {\bibfnamefont {T.}~\bibnamefont {Tsegaye}},\
  and\ \bibinfo {author} {\bibfnamefont {A.}~\bibnamefont {Trifonov}},\
  }\bibfield  {title} {\bibinfo {title} {States that minimize the evolution
  time to become an orthogonal state},\ }\href
  {https://doi.org/10.1103/PhysRevA.59.1788} {\bibfield  {journal} {\bibinfo
  {journal} {Physical Review A}\ }\textbf {\bibinfo {volume} {59}},\ \bibinfo
  {pages} {1788} (\bibinfo {year} {1999})}\BibitemShut {NoStop}%
\bibitem [{\citenamefont {Andrecut}\ and\ \citenamefont
  {Ali}(2004)}]{Andrecut2004}%
  \BibitemOpen
  \bibfield  {author} {\bibinfo {author} {\bibfnamefont {M.}~\bibnamefont
  {Andrecut}}\ and\ \bibinfo {author} {\bibfnamefont {M.~K.}\ \bibnamefont
  {Ali}},\ }\bibfield  {title} {\bibinfo {title} {The adiabatic analogue of the
  {M}argolus{\textendash}{L}evitin theorem},\ }\href
  {https://doi.org/10.1088/0305-4470/37/15/l01} {\bibfield  {journal} {\bibinfo
   {journal} {Journal of Physics A: Mathematical and General}\ }\textbf
  {\bibinfo {volume} {37}},\ \bibinfo {pages} {L157} (\bibinfo {year}
  {2004})}\BibitemShut {NoStop}%
\bibitem [{\citenamefont {Gray}\ and\ \citenamefont {Vogt}(2005)}]{Gray2005}%
  \BibitemOpen
  \bibfield  {author} {\bibinfo {author} {\bibfnamefont {J.~E.}\ \bibnamefont
  {Gray}}\ and\ \bibinfo {author} {\bibfnamefont {A.}~\bibnamefont {Vogt}},\
  }\bibfield  {title} {\bibinfo {title} {Mathematical analysis of the
  {M}andelstam--{T}amm time-energy uncertainty principle},\ }\href
  {https://doi.org/10.1063/1.1897164} {\bibfield  {journal} {\bibinfo
  {journal} {Journal of mathematical physics}\ }\textbf {\bibinfo {volume}
  {46}},\ \bibinfo {pages} {052108} (\bibinfo {year} {2005})}\BibitemShut
  {NoStop}%
\bibitem [{\citenamefont {Luo}\ and\ \citenamefont {Zhang}(2005)}]{Luo2005}%
  \BibitemOpen
  \bibfield  {author} {\bibinfo {author} {\bibfnamefont {S.}~\bibnamefont
  {Luo}}\ and\ \bibinfo {author} {\bibfnamefont {Z.}~\bibnamefont {Zhang}},\
  }\bibfield  {title} {\bibinfo {title} {On {D}ecaying {R}ate of {Q}uantum
  {S}tates},\ }\href {https://doi.org/10.1007/s11005-004-5095-4} {\bibfield
  {journal} {\bibinfo  {journal} {Letters in Mathematical Physics}\ }\textbf
  {\bibinfo {volume} {71}},\ \bibinfo {pages} {1} (\bibinfo {year}
  {2005})}\BibitemShut {NoStop}%
\bibitem [{\citenamefont {Zieli{\'n}ski}\ and\ \citenamefont
  {Zych}(2006)}]{Zielinski2006}%
  \BibitemOpen
  \bibfield  {author} {\bibinfo {author} {\bibfnamefont {B.}~\bibnamefont
  {Zieli{\'n}ski}}\ and\ \bibinfo {author} {\bibfnamefont {M.}~\bibnamefont
  {Zych}},\ }\bibfield  {title} {\bibinfo {title} {Generalization of the
  {M}argolus-{L}evitin bound},\ }\href
  {https://doi.org/10.1103/PhysRevA.74.034301} {\bibfield  {journal} {\bibinfo
  {journal} {Physical Review A}\ }\textbf {\bibinfo {volume} {74}},\ \bibinfo
  {pages} {034301} (\bibinfo {year} {2006})}\BibitemShut {NoStop}%
\bibitem [{\citenamefont {Andrews}(2007)}]{Andrews2007}%
  \BibitemOpen
  \bibfield  {author} {\bibinfo {author} {\bibfnamefont {M.}~\bibnamefont
  {Andrews}},\ }\bibfield  {title} {\bibinfo {title} {Bounds to unitary
  evolution},\ }\href {https://doi.org/10.1103/PhysRevA.75.062112} {\bibfield
  {journal} {\bibinfo  {journal} {Physical Review A}\ }\textbf {\bibinfo
  {volume} {75}},\ \bibinfo {pages} {062112} (\bibinfo {year}
  {2007})}\BibitemShut {NoStop}%
\bibitem [{\citenamefont {Yurtsever}(2010)}]{Yurtsever2010}%
  \BibitemOpen
  \bibfield  {author} {\bibinfo {author} {\bibfnamefont {U.}~\bibnamefont
  {Yurtsever}},\ }\bibfield  {title} {\bibinfo {title} {Fundamental limits on
  the speed of evolution of quantum states},\ }\href
  {https://doi.org/10.1088/0031-8949/82/03/035008} {\bibfield  {journal}
  {\bibinfo  {journal} {Physica Scripta}\ }\textbf {\bibinfo {volume} {82}},\
  \bibinfo {pages} {035008} (\bibinfo {year} {2010})}\BibitemShut {NoStop}%
\bibitem [{\citenamefont {Shuang-Shuang}\ \emph {et~al.}(2010)\citenamefont
  {Shuang-Shuang}, \citenamefont {Nan},\ and\ \citenamefont
  {Shun-Long}}]{Fu2010}%
  \BibitemOpen
  \bibfield  {author} {\bibinfo {author} {\bibfnamefont {F.}~\bibnamefont
  {Shuang-Shuang}}, \bibinfo {author} {\bibfnamefont {L.}~\bibnamefont {Nan}},\
  and\ \bibinfo {author} {\bibfnamefont {L.}~\bibnamefont {Shun-Long}},\
  }\bibfield  {title} {\bibinfo {title} {A {N}ote on {F}undamental {L}imit of
  {Q}uantum {D}ynamics {R}ate},\ }\href
  {https://doi.org/10.1088/0253-6102/54/4/15} {\bibfield  {journal} {\bibinfo
  {journal} {Communications in Theoretical Physics}\ }\textbf {\bibinfo
  {volume} {54}},\ \bibinfo {pages} {661} (\bibinfo {year} {2010})}\BibitemShut
  {NoStop}%
\bibitem [{\citenamefont {Zwierz}(2012)}]{Zwierz2012}%
  \BibitemOpen
  \bibfield  {author} {\bibinfo {author} {\bibfnamefont {M.}~\bibnamefont
  {Zwierz}},\ }\bibfield  {title} {\bibinfo {title} {Comment on ``{G}eometric
  derivation of the quantum speed limit''},\ }\href
  {https://doi.org/10.1103/PhysRevA.86.016101} {\bibfield  {journal} {\bibinfo
  {journal} {Physical Review A}\ }\textbf {\bibinfo {volume} {86}},\ \bibinfo
  {pages} {016101} (\bibinfo {year} {2012})}\BibitemShut {NoStop}%
\bibitem [{\citenamefont {Poggi}\ \emph {et~al.}(2013)\citenamefont {Poggi},
  \citenamefont {Lombardo},\ and\ \citenamefont {Wisniacki}}]{Poggi2013}%
  \BibitemOpen
  \bibfield  {author} {\bibinfo {author} {\bibfnamefont {P.~M.}\ \bibnamefont
  {Poggi}}, \bibinfo {author} {\bibfnamefont {F.~C.}\ \bibnamefont
  {Lombardo}},\ and\ \bibinfo {author} {\bibfnamefont {D.~A.}\ \bibnamefont
  {Wisniacki}},\ }\bibfield  {title} {\bibinfo {title} {Quantum speed limit and
  optimal evolution time in a two-level system},\ }\href
  {https://doi.org/10.1209/0295-5075/104/40005} {\bibfield  {journal} {\bibinfo
   {journal} {Europhysics Letters (EPL)}\ }\textbf {\bibinfo {volume} {104}},\
  \bibinfo {pages} {40005} (\bibinfo {year} {2013})}\BibitemShut {NoStop}%
\bibitem [{\citenamefont {Kupferman}\ and\ \citenamefont
  {Reznik}(2008)}]{Kupferman2008}%
  \BibitemOpen
  \bibfield  {author} {\bibinfo {author} {\bibfnamefont {J.}~\bibnamefont
  {Kupferman}}\ and\ \bibinfo {author} {\bibfnamefont {B.}~\bibnamefont
  {Reznik}},\ }\bibfield  {title} {\bibinfo {title} {Entanglement and the speed
  of evolution in mixed states},\ }\href
  {https://doi.org/10.1103/PhysRevA.78.042305} {\bibfield  {journal} {\bibinfo
  {journal} {Physical Review A}\ }\textbf {\bibinfo {volume} {78}},\ \bibinfo
  {pages} {042305} (\bibinfo {year} {2008})}\BibitemShut {NoStop}%
\bibitem [{\citenamefont {Jones}\ and\ \citenamefont {Kok}(2010)}]{Jones2010}%
  \BibitemOpen
  \bibfield  {author} {\bibinfo {author} {\bibfnamefont {P.~J.}\ \bibnamefont
  {Jones}}\ and\ \bibinfo {author} {\bibfnamefont {P.}~\bibnamefont {Kok}},\
  }\bibfield  {title} {\bibinfo {title} {Geometric derivation of the quantum
  speed limit},\ }\href {https://doi.org/10.1103/PhysRevA.82.022107} {\bibfield
   {journal} {\bibinfo  {journal} {Physical Review A}\ }\textbf {\bibinfo
  {volume} {82}},\ \bibinfo {pages} {022107} (\bibinfo {year}
  {2010})}\BibitemShut {NoStop}%
\bibitem [{\citenamefont {Chau}(2010)}]{Chau2010}%
  \BibitemOpen
  \bibfield  {author} {\bibinfo {author} {\bibfnamefont {H.~F.}\ \bibnamefont
  {Chau}},\ }\bibfield  {title} {\bibinfo {title} {Tight upper bound of the
  maximum speed of evolution of a quantum state},\ }\href
  {https://doi.org/10.1103/PhysRevA.81.062133} {\bibfield  {journal} {\bibinfo
  {journal} {Physical Review A}\ }\textbf {\bibinfo {volume} {81}},\ \bibinfo
  {pages} {062133} (\bibinfo {year} {2010})}\BibitemShut {NoStop}%
\bibitem [{\citenamefont {Deffner}\ and\ \citenamefont
  {Lutz}(2013{\natexlab{a}})}]{S.Deffner2013}%
  \BibitemOpen
  \bibfield  {author} {\bibinfo {author} {\bibfnamefont {S.}~\bibnamefont
  {Deffner}}\ and\ \bibinfo {author} {\bibfnamefont {E.}~\bibnamefont {Lutz}},\
  }\bibfield  {title} {\bibinfo {title} {Energy{\textendash}time uncertainty
  relation for driven quantum systems},\ }\href
  {https://doi.org/10.1088/1751-8113/46/33/335302} {\bibfield  {journal}
  {\bibinfo  {journal} {Journal of Physics A: Mathematical and Theoretical}\
  }\textbf {\bibinfo {volume} {46}},\ \bibinfo {pages} {335302} (\bibinfo
  {year} {2013}{\natexlab{a}})}\BibitemShut {NoStop}%
\bibitem [{\citenamefont {Fung}\ and\ \citenamefont {Chau}(2014)}]{Fung2014}%
  \BibitemOpen
  \bibfield  {author} {\bibinfo {author} {\bibfnamefont {C.-H.~F.}\
  \bibnamefont {Fung}}\ and\ \bibinfo {author} {\bibfnamefont {H.}~\bibnamefont
  {Chau}},\ }\bibfield  {title} {\bibinfo {title} {Relation between physical
  time-energy cost of a quantum process and its information fidelity},\ }\href
  {https://doi.org/10.1103/PhysRevA.90.022333} {\bibfield  {journal} {\bibinfo
  {journal} {Physical Review A}\ }\textbf {\bibinfo {volume} {90}},\ \bibinfo
  {pages} {022333} (\bibinfo {year} {2014})}\BibitemShut {NoStop}%
\bibitem [{\citenamefont {Andersson}\ and\ \citenamefont
  {Heydari}(2014)}]{Andersson2014}%
  \BibitemOpen
  \bibfield  {author} {\bibinfo {author} {\bibfnamefont {O.}~\bibnamefont
  {Andersson}}\ and\ \bibinfo {author} {\bibfnamefont {H.}~\bibnamefont
  {Heydari}},\ }\bibfield  {title} {\bibinfo {title} {Quantum speed limits and
  optimal {H}amiltonians for driven systems in mixed states},\ }\href
  {https://doi.org/10.1088/1751-8113/47/21/215301} {\bibfield  {journal}
  {\bibinfo  {journal} {Journal of Physics A: Mathematical and Theoretical}\
  }\textbf {\bibinfo {volume} {47}},\ \bibinfo {pages} {215301} (\bibinfo
  {year} {2014})}\BibitemShut {NoStop}%
\bibitem [{\citenamefont {Mondal}\ \emph {et~al.}(2016)\citenamefont {Mondal},
  \citenamefont {Datta},\ and\ \citenamefont {Sazim}}]{D.Mondal2016}%
  \BibitemOpen
  \bibfield  {author} {\bibinfo {author} {\bibfnamefont {D.}~\bibnamefont
  {Mondal}}, \bibinfo {author} {\bibfnamefont {C.}~\bibnamefont {Datta}},\ and\
  \bibinfo {author} {\bibfnamefont {S.}~\bibnamefont {Sazim}},\ }\bibfield
  {title} {\bibinfo {title} {Quantum coherence sets the quantum speed limit for
  mixed states},\ }\href
  {https://doi.org/https://doi.org/10.1016/j.physleta.2015.12.015} {\bibfield
  {journal} {\bibinfo  {journal} {Physics Letters A}\ }\textbf {\bibinfo
  {volume} {380}},\ \bibinfo {pages} {689} (\bibinfo {year}
  {2016})}\BibitemShut {NoStop}%
\bibitem [{\citenamefont {Mondal}\ and\ \citenamefont
  {Pati}(2016)}]{Mondal2016}%
  \BibitemOpen
  \bibfield  {author} {\bibinfo {author} {\bibfnamefont {D.}~\bibnamefont
  {Mondal}}\ and\ \bibinfo {author} {\bibfnamefont {A.~K.}\ \bibnamefont
  {Pati}},\ }\bibfield  {title} {\bibinfo {title} {Quantum speed limit for
  mixed states using an experimentally realizable metric},\ }\href
  {https://doi.org/https://doi.org/10.1016/j.physleta.2016.02.018} {\bibfield
  {journal} {\bibinfo  {journal} {Physics Letters A}\ }\textbf {\bibinfo
  {volume} {380}},\ \bibinfo {pages} {1395} (\bibinfo {year}
  {2016})}\BibitemShut {NoStop}%
\bibitem [{\citenamefont {Deffner}\ and\ \citenamefont
  {Campbell}(2017)}]{S.Deffner2017}%
  \BibitemOpen
  \bibfield  {author} {\bibinfo {author} {\bibfnamefont {S.}~\bibnamefont
  {Deffner}}\ and\ \bibinfo {author} {\bibfnamefont {S.}~\bibnamefont
  {Campbell}},\ }\bibfield  {title} {\bibinfo {title} {Quantum speed limits:
  from {H}eisenberg's uncertainty principle to optimal quantum control},\
  }\href {https://doi.org/10.1088/1751-8121/aa86c6} {\bibfield  {journal}
  {\bibinfo  {journal} {Journal of Physics A: Mathematical and Theoretical}\
  }\textbf {\bibinfo {volume} {50}},\ \bibinfo {pages} {453001} (\bibinfo
  {year} {2017})}\BibitemShut {NoStop}%
\bibitem [{\citenamefont {Campaioli}\ \emph {et~al.}(2018)\citenamefont
  {Campaioli}, \citenamefont {Pollock}, \citenamefont {Binder},\ and\
  \citenamefont {Modi}}]{Campaioli2018}%
  \BibitemOpen
  \bibfield  {author} {\bibinfo {author} {\bibfnamefont {F.}~\bibnamefont
  {Campaioli}}, \bibinfo {author} {\bibfnamefont {F.~A.}\ \bibnamefont
  {Pollock}}, \bibinfo {author} {\bibfnamefont {F.~C.}\ \bibnamefont
  {Binder}},\ and\ \bibinfo {author} {\bibfnamefont {K.}~\bibnamefont {Modi}},\
  }\bibfield  {title} {\bibinfo {title} {Tightening {Q}uantum {S}peed {L}imits
  for {A}lmost {A}ll {S}tates},\ }\href
  {https://doi.org/10.1103/PhysRevLett.120.060409} {\bibfield  {journal}
  {\bibinfo  {journal} {Physical Review Letters}\ }\textbf {\bibinfo {volume}
  {120}},\ \bibinfo {pages} {060409} (\bibinfo {year} {2018})}\BibitemShut
  {NoStop}%
\bibitem [{\citenamefont {Giovannetti}\ \emph {et~al.}(2004)\citenamefont
  {Giovannetti}, \citenamefont {Lloyd},\ and\ \citenamefont
  {Maccone}}]{Giovannetti2004}%
  \BibitemOpen
  \bibfield  {author} {\bibinfo {author} {\bibfnamefont {V.}~\bibnamefont
  {Giovannetti}}, \bibinfo {author} {\bibfnamefont {S.}~\bibnamefont {Lloyd}},\
  and\ \bibinfo {author} {\bibfnamefont {L.}~\bibnamefont {Maccone}},\
  }\bibfield  {title} {\bibinfo {title} {The speed limit of quantum unitary
  evolution},\ }\href {https://doi.org/10.1088/1464-4266/6/8/028} {\bibfield
  {journal} {\bibinfo  {journal} {Journal of Optics B: Quantum and
  Semiclassical Optics}\ }\textbf {\bibinfo {volume} {6}},\ \bibinfo {pages}
  {S807} (\bibinfo {year} {2004})}\BibitemShut {NoStop}%
\bibitem [{\citenamefont {Batle}\ \emph {et~al.}(2005)\citenamefont {Batle},
  \citenamefont {Casas}, \citenamefont {Plastino},\ and\ \citenamefont
  {Plastino}}]{Batle2005}%
  \BibitemOpen
  \bibfield  {author} {\bibinfo {author} {\bibfnamefont {J.}~\bibnamefont
  {Batle}}, \bibinfo {author} {\bibfnamefont {M.}~\bibnamefont {Casas}},
  \bibinfo {author} {\bibfnamefont {A.}~\bibnamefont {Plastino}},\ and\
  \bibinfo {author} {\bibfnamefont {A.~R.}\ \bibnamefont {Plastino}},\
  }\bibfield  {title} {\bibinfo {title} {Connection between entanglement and
  the speed of quantum evolution},\ }\href
  {https://doi.org/10.1103/PhysRevA.72.032337} {\bibfield  {journal} {\bibinfo
  {journal} {Physical Review A}\ }\textbf {\bibinfo {volume} {72}},\ \bibinfo
  {pages} {032337} (\bibinfo {year} {2005})}\BibitemShut {NoStop}%
\bibitem [{\citenamefont {Borr\'as}\ \emph {et~al.}(2006)\citenamefont
  {Borr\'as}, \citenamefont {Casas}, \citenamefont {Plastino},\ and\
  \citenamefont {Plastino}}]{Borras2006}%
  \BibitemOpen
  \bibfield  {author} {\bibinfo {author} {\bibfnamefont {A.}~\bibnamefont
  {Borr\'as}}, \bibinfo {author} {\bibfnamefont {M.}~\bibnamefont {Casas}},
  \bibinfo {author} {\bibfnamefont {A.~R.}\ \bibnamefont {Plastino}},\ and\
  \bibinfo {author} {\bibfnamefont {A.}~\bibnamefont {Plastino}},\ }\bibfield
  {title} {\bibinfo {title} {Entanglement and the lower bounds on the speed of
  quantum evolution},\ }\href {https://doi.org/10.1103/PhysRevA.74.022326}
  {\bibfield  {journal} {\bibinfo  {journal} {Physical Review A}\ }\textbf
  {\bibinfo {volume} {74}},\ \bibinfo {pages} {022326} (\bibinfo {year}
  {2006})}\BibitemShut {NoStop}%
\bibitem [{\citenamefont {Zander}\ \emph {et~al.}(2007)\citenamefont {Zander},
  \citenamefont {Plastino}, \citenamefont {Plastino},\ and\ \citenamefont
  {Casas}}]{Zander2007}%
  \BibitemOpen
  \bibfield  {author} {\bibinfo {author} {\bibfnamefont {C.}~\bibnamefont
  {Zander}}, \bibinfo {author} {\bibfnamefont {A.~R.}\ \bibnamefont
  {Plastino}}, \bibinfo {author} {\bibfnamefont {A.}~\bibnamefont {Plastino}},\
  and\ \bibinfo {author} {\bibfnamefont {M.}~\bibnamefont {Casas}},\ }\bibfield
   {title} {\bibinfo {title} {Entanglement and the speed of evolution of
  multi-partite quantum systems},\ }\href
  {https://doi.org/10.1088/1751-8113/40/11/020} {\bibfield  {journal} {\bibinfo
   {journal} {Journal of Physics A: Mathematical and Theoretical}\ }\textbf
  {\bibinfo {volume} {40}},\ \bibinfo {pages} {2861} (\bibinfo {year}
  {2007})}\BibitemShut {NoStop}%
\bibitem [{\citenamefont {Ness}\ \emph {et~al.}(2022)\citenamefont {Ness},
  \citenamefont {Alberti},\ and\ \citenamefont {Sagi}}]{Ness2022}%
  \BibitemOpen
  \bibfield  {author} {\bibinfo {author} {\bibfnamefont {G.}~\bibnamefont
  {Ness}}, \bibinfo {author} {\bibfnamefont {A.}~\bibnamefont {Alberti}},\ and\
  \bibinfo {author} {\bibfnamefont {Y.}~\bibnamefont {Sagi}},\ }\bibfield
  {title} {\bibinfo {title} {Quantum speed limit for states with a bounded
  energy spectrum},\ }\href {https://doi.org/10.1103/PhysRevLett.129.140403}
  {\bibfield  {journal} {\bibinfo  {journal} {Phys. Rev. Lett.}\ }\textbf
  {\bibinfo {volume} {129}},\ \bibinfo {pages} {140403} (\bibinfo {year}
  {2022})}\BibitemShut {NoStop}%
\bibitem [{\citenamefont {Shrimali}\ \emph {et~al.}(2022)\citenamefont
  {Shrimali}, \citenamefont {Bhowmick}, \citenamefont {Pandey},\ and\
  \citenamefont {Pati}}]{Pandey22}%
  \BibitemOpen
  \bibfield  {author} {\bibinfo {author} {\bibfnamefont {D.}~\bibnamefont
  {Shrimali}}, \bibinfo {author} {\bibfnamefont {S.}~\bibnamefont {Bhowmick}},
  \bibinfo {author} {\bibfnamefont {V.}~\bibnamefont {Pandey}},\ and\ \bibinfo
  {author} {\bibfnamefont {A.~K.}\ \bibnamefont {Pati}},\ }\bibfield  {title}
  {\bibinfo {title} {Capacity of entanglement for a nonlocal hamiltonian},\
  }\href {https://doi.org/10.1103/PhysRevA.106.042419} {\bibfield  {journal}
  {\bibinfo  {journal} {Phys. Rev. A}\ }\textbf {\bibinfo {volume} {106}},\
  \bibinfo {pages} {042419} (\bibinfo {year} {2022})}\BibitemShut {NoStop}%
\bibitem [{\citenamefont {Thakuria}\ and\ \citenamefont
  {Pati}(2022)}]{thakuria2022}%
  \BibitemOpen
  \bibfield  {author} {\bibinfo {author} {\bibfnamefont {D.}~\bibnamefont
  {Thakuria}}\ and\ \bibinfo {author} {\bibfnamefont {A.~K.}\ \bibnamefont
  {Pati}},\ }\bibfield  {title} {\bibinfo {title} {Stronger quantum speed
  limit},\ }\href {https://doi.org/10.48550/arxiv.2208.05469} {\bibfield
  {journal} {\bibinfo  {journal} {arXiv:2208.05469}\ } (\bibinfo {year}
  {2022})}\BibitemShut {NoStop}%
\bibitem [{\citenamefont {Deffner}\ and\ \citenamefont
  {Lutz}(2013{\natexlab{b}})}]{Deffner2013}%
  \BibitemOpen
  \bibfield  {author} {\bibinfo {author} {\bibfnamefont {S.}~\bibnamefont
  {Deffner}}\ and\ \bibinfo {author} {\bibfnamefont {E.}~\bibnamefont {Lutz}},\
  }\bibfield  {title} {\bibinfo {title} {Quantum {S}peed {L}imit for
  {N}on-{M}arkovian {D}ynamics},\ }\href
  {https://doi.org/10.1103/PhysRevLett.111.010402} {\bibfield  {journal}
  {\bibinfo  {journal} {Physical Review Letters}\ }\textbf {\bibinfo {volume}
  {111}},\ \bibinfo {pages} {010402} (\bibinfo {year}
  {2013}{\natexlab{b}})}\BibitemShut {NoStop}%
\bibitem [{\citenamefont {del Campo}\ \emph {et~al.}(2013)\citenamefont {del
  Campo}, \citenamefont {Egusquiza}, \citenamefont {Plenio},\ and\
  \citenamefont {Huelga}}]{Campo2013}%
  \BibitemOpen
  \bibfield  {author} {\bibinfo {author} {\bibfnamefont {A.}~\bibnamefont {del
  Campo}}, \bibinfo {author} {\bibfnamefont {I.~L.}\ \bibnamefont {Egusquiza}},
  \bibinfo {author} {\bibfnamefont {M.~B.}\ \bibnamefont {Plenio}},\ and\
  \bibinfo {author} {\bibfnamefont {S.~F.}\ \bibnamefont {Huelga}},\ }\bibfield
   {title} {\bibinfo {title} {Quantum {S}peed {L}imits in {O}pen {S}ystem
  {D}ynamics},\ }\href {https://doi.org/10.1103/PhysRevLett.110.050403}
  {\bibfield  {journal} {\bibinfo  {journal} {Physical Review Letters}\
  }\textbf {\bibinfo {volume} {110}},\ \bibinfo {pages} {050403} (\bibinfo
  {year} {2013})}\BibitemShut {NoStop}%
\bibitem [{\citenamefont {Taddei}\ \emph {et~al.}(2013)\citenamefont {Taddei},
  \citenamefont {Escher}, \citenamefont {Davidovich},\ and\ \citenamefont
  {de~Matos~Filho}}]{Taddei2013}%
  \BibitemOpen
  \bibfield  {author} {\bibinfo {author} {\bibfnamefont {M.~M.}\ \bibnamefont
  {Taddei}}, \bibinfo {author} {\bibfnamefont {B.~M.}\ \bibnamefont {Escher}},
  \bibinfo {author} {\bibfnamefont {L.}~\bibnamefont {Davidovich}},\ and\
  \bibinfo {author} {\bibfnamefont {R.~L.}\ \bibnamefont {de~Matos~Filho}},\
  }\bibfield  {title} {\bibinfo {title} {Quantum {S}peed {L}imit for {P}hysical
  {P}rocesses},\ }\href {https://doi.org/10.1103/PhysRevLett.110.050402}
  {\bibfield  {journal} {\bibinfo  {journal} {Physical Review Letters}\
  }\textbf {\bibinfo {volume} {110}},\ \bibinfo {pages} {050402} (\bibinfo
  {year} {2013})}\BibitemShut {NoStop}%
\bibitem [{\citenamefont {Fung}\ and\ \citenamefont {Chau}(2013)}]{Fung2013}%
  \BibitemOpen
  \bibfield  {author} {\bibinfo {author} {\bibfnamefont {C.-H.~F.}\
  \bibnamefont {Fung}}\ and\ \bibinfo {author} {\bibfnamefont {H.~F.}\
  \bibnamefont {Chau}},\ }\bibfield  {title} {\bibinfo {title} {Time-energy
  measure for quantum processes},\ }\href
  {https://doi.org/10.1103/PhysRevA.88.012307} {\bibfield  {journal} {\bibinfo
  {journal} {Physical Review A}\ }\textbf {\bibinfo {volume} {88}},\ \bibinfo
  {pages} {012307} (\bibinfo {year} {2013})}\BibitemShut {NoStop}%
\bibitem [{\citenamefont {Pires}\ \emph {et~al.}(2016)\citenamefont {Pires},
  \citenamefont {Cianciaruso}, \citenamefont {C\'eleri}, \citenamefont
  {Adesso},\ and\ \citenamefont {Soares-Pinto}}]{Pires2016}%
  \BibitemOpen
  \bibfield  {author} {\bibinfo {author} {\bibfnamefont {D.~P.}\ \bibnamefont
  {Pires}}, \bibinfo {author} {\bibfnamefont {M.}~\bibnamefont {Cianciaruso}},
  \bibinfo {author} {\bibfnamefont {L.~C.}\ \bibnamefont {C\'eleri}}, \bibinfo
  {author} {\bibfnamefont {G.}~\bibnamefont {Adesso}},\ and\ \bibinfo {author}
  {\bibfnamefont {D.~O.}\ \bibnamefont {Soares-Pinto}},\ }\bibfield  {title}
  {\bibinfo {title} {Generalized {G}eometric {Q}uantum {S}peed {L}imits},\
  }\href {https://doi.org/10.1103/PhysRevX.6.021031} {\bibfield  {journal}
  {\bibinfo  {journal} {Physical Review X}\ }\textbf {\bibinfo {volume} {6}},\
  \bibinfo {pages} {021031} (\bibinfo {year} {2016})}\BibitemShut {NoStop}%
\bibitem [{\citenamefont {Deffner}(2020)}]{S.Deffner2020}%
  \BibitemOpen
  \bibfield  {author} {\bibinfo {author} {\bibfnamefont {S.}~\bibnamefont
  {Deffner}},\ }\bibfield  {title} {\bibinfo {title} {Quantum speed limits and
  the maximal rate of information production},\ }\href
  {https://doi.org/10.1103/PhysRevResearch.2.013161} {\bibfield  {journal}
  {\bibinfo  {journal} {Physical Review Research}\ }\textbf {\bibinfo {volume}
  {2}},\ \bibinfo {pages} {013161} (\bibinfo {year} {2020})}\BibitemShut
  {NoStop}%
\bibitem [{\citenamefont {Jing}\ \emph {et~al.}(2016)\citenamefont {Jing},
  \citenamefont {Wu},\ and\ \citenamefont {Del~Campo}}]{Jing2016}%
  \BibitemOpen
  \bibfield  {author} {\bibinfo {author} {\bibfnamefont {J.}~\bibnamefont
  {Jing}}, \bibinfo {author} {\bibfnamefont {L.-A.}\ \bibnamefont {Wu}},\ and\
  \bibinfo {author} {\bibfnamefont {A.}~\bibnamefont {Del~Campo}},\ }\bibfield
  {title} {\bibinfo {title} {Fundamental {S}peed {L}imits to the {G}eneration
  of {Q}uantumness},\ }\href {https://doi.org/10.1038/srep38149} {\bibfield
  {journal} {\bibinfo  {journal} {Scientific Reports}\ }\textbf {\bibinfo
  {volume} {6}},\ \bibinfo {pages} {38149} (\bibinfo {year}
  {2016})}\BibitemShut {NoStop}%
\bibitem [{\citenamefont {Garc\'{\i}a-Pintos}\ \emph
  {et~al.}(2022)\citenamefont {Garc\'{\i}a-Pintos}, \citenamefont {Nicholson},
  \citenamefont {Green}, \citenamefont {del Campo},\ and\ \citenamefont
  {Gorshkov}}]{Pintos2021}%
  \BibitemOpen
  \bibfield  {author} {\bibinfo {author} {\bibfnamefont {L.~P.}\ \bibnamefont
  {Garc\'{\i}a-Pintos}}, \bibinfo {author} {\bibfnamefont {S.~B.}\ \bibnamefont
  {Nicholson}}, \bibinfo {author} {\bibfnamefont {J.~R.}\ \bibnamefont
  {Green}}, \bibinfo {author} {\bibfnamefont {A.}~\bibnamefont {del Campo}},\
  and\ \bibinfo {author} {\bibfnamefont {A.~V.}\ \bibnamefont {Gorshkov}},\
  }\bibfield  {title} {\bibinfo {title} {Unifying {Q}uantum and {C}lassical
  {S}peed {L}imits on {O}bservables},\ }\href
  {https://doi.org/10.1103/PhysRevX.12.011038} {\bibfield  {journal} {\bibinfo
  {journal} {Physical Review X}\ }\textbf {\bibinfo {volume} {12}},\ \bibinfo
  {pages} {011038} (\bibinfo {year} {2022})}\BibitemShut {NoStop}%
\bibitem [{\citenamefont {Mohan}\ \emph {et~al.}(2022)\citenamefont {Mohan},
  \citenamefont {Das},\ and\ \citenamefont {Pati}}]{Mohan}%
  \BibitemOpen
  \bibfield  {author} {\bibinfo {author} {\bibfnamefont {B.}~\bibnamefont
  {Mohan}}, \bibinfo {author} {\bibfnamefont {S.}~\bibnamefont {Das}},\ and\
  \bibinfo {author} {\bibfnamefont {A.~K.}\ \bibnamefont {Pati}},\ }\bibfield
  {title} {\bibinfo {title} {Quantum speed limits for information and
  coherence},\ }\href {https://doi.org/10.1088/1367-2630/ac753c} {\bibfield
  {journal} {\bibinfo  {journal} {New Journal of Physics}\ }\textbf {\bibinfo
  {volume} {24}},\ \bibinfo {pages} {065003} (\bibinfo {year}
  {2022})}\BibitemShut {NoStop}%
\bibitem [{\citenamefont {Mohan}\ and\ \citenamefont {Pati}(2022)}]{Mohan21}%
  \BibitemOpen
  \bibfield  {author} {\bibinfo {author} {\bibfnamefont {B.}~\bibnamefont
  {Mohan}}\ and\ \bibinfo {author} {\bibfnamefont {A.~K.}\ \bibnamefont
  {Pati}},\ }\bibfield  {title} {\bibinfo {title} {Quantum speed limits for
  observables},\ }\href {https://doi.org/10.1103/PhysRevA.106.042436}
  {\bibfield  {journal} {\bibinfo  {journal} {Phys. Rev. A}\ }\textbf {\bibinfo
  {volume} {106}},\ \bibinfo {pages} {042436} (\bibinfo {year}
  {2022})}\BibitemShut {NoStop}%
\bibitem [{\citenamefont {Pandey}\ \emph {et~al.}(2022)\citenamefont {Pandey},
  \citenamefont {Shrimali}, \citenamefont {Mohan}, \citenamefont {Das},\ and\
  \citenamefont {Pati}}]{Pandey2022}%
  \BibitemOpen
  \bibfield  {author} {\bibinfo {author} {\bibfnamefont {V.}~\bibnamefont
  {Pandey}}, \bibinfo {author} {\bibfnamefont {D.}~\bibnamefont {Shrimali}},
  \bibinfo {author} {\bibfnamefont {B.}~\bibnamefont {Mohan}}, \bibinfo
  {author} {\bibfnamefont {S.}~\bibnamefont {Das}},\ and\ \bibinfo {author}
  {\bibfnamefont {A.~K.}\ \bibnamefont {Pati}},\ }\bibfield  {title} {\bibinfo
  {title} {Speed limits on correlations in bipartite quantum systems},\ }\href
  {https://doi.org/10.48550/arXiv.2207.05645} {\bibfield  {journal} {\bibinfo
  {journal} {arXiv preprint arXiv:2207.05645}\ } (\bibinfo {year}
  {2022})}\BibitemShut {NoStop}%
\bibitem [{\citenamefont {Thakuria}\ \emph {et~al.}(2022)\citenamefont
  {Thakuria}, \citenamefont {Srivastav}, \citenamefont {Mohan}, \citenamefont
  {Kumari},\ and\ \citenamefont {Pati}}]{abhay2022}%
  \BibitemOpen
  \bibfield  {author} {\bibinfo {author} {\bibfnamefont {D.}~\bibnamefont
  {Thakuria}}, \bibinfo {author} {\bibfnamefont {A.}~\bibnamefont {Srivastav}},
  \bibinfo {author} {\bibfnamefont {B.}~\bibnamefont {Mohan}}, \bibinfo
  {author} {\bibfnamefont {A.}~\bibnamefont {Kumari}},\ and\ \bibinfo {author}
  {\bibfnamefont {A.~K.}\ \bibnamefont {Pati}},\ }\bibfield  {title} {\bibinfo
  {title} {Generalised quantum speed limit for arbitrary evolution},\ }\href
  {https://doi.org/10.48550/arxiv.2207.04124} {\bibfield  {journal} {\bibinfo
  {journal} {arXiv:2207.04124}\ } (\bibinfo {year} {2022})}\BibitemShut
  {NoStop}%
\bibitem [{\citenamefont {Carabba}\ \emph {et~al.}(2022)\citenamefont
  {Carabba}, \citenamefont {H{\"{o}}rnedal},\ and\ \citenamefont
  {Campo}}]{Carabba2022}%
  \BibitemOpen
  \bibfield  {author} {\bibinfo {author} {\bibfnamefont {N.}~\bibnamefont
  {Carabba}}, \bibinfo {author} {\bibfnamefont {N.}~\bibnamefont
  {H{\"{o}}rnedal}},\ and\ \bibinfo {author} {\bibfnamefont {A.~d.}\
  \bibnamefont {Campo}},\ }\bibfield  {title} {\bibinfo {title} {Quantum speed
  limits on operator flows and correlation functions},\ }\href
  {https://doi.org/10.22331/q-2022-12-22-884} {\bibfield  {journal} {\bibinfo
  {journal} {{Quantum}}\ }\textbf {\bibinfo {volume} {6}},\ \bibinfo {pages}
  {884} (\bibinfo {year} {2022})}\BibitemShut {NoStop}%
\bibitem [{\citenamefont {Naseri}\ \emph {et~al.}(2022)\citenamefont {Naseri},
  \citenamefont {Macchiavello}, \citenamefont {Bruß}, \citenamefont
  {Horodecki},\ and\ \citenamefont {Streltsov}}]{naseri2022}%
  \BibitemOpen
  \bibfield  {author} {\bibinfo {author} {\bibfnamefont {M.}~\bibnamefont
  {Naseri}}, \bibinfo {author} {\bibfnamefont {C.}~\bibnamefont
  {Macchiavello}}, \bibinfo {author} {\bibfnamefont {D.}~\bibnamefont {Bruß}},
  \bibinfo {author} {\bibfnamefont {P.}~\bibnamefont {Horodecki}},\ and\
  \bibinfo {author} {\bibfnamefont {A.}~\bibnamefont {Streltsov}},\ }\bibfield
  {title} {\bibinfo {title} {Quantum speed limit for change of basis},\ }\href
  {https://doi.org/10.48550/arxiv.2212.12352} {\bibfield  {journal} {\bibinfo
  {journal} {arXiv:2212.12352}\ } (\bibinfo {year} {2022})}\BibitemShut
  {NoStop}%
\bibitem [{\citenamefont {Meng}\ and\ \citenamefont {Xu}(2022)}]{meng2022}%
  \BibitemOpen
  \bibfield  {author} {\bibinfo {author} {\bibfnamefont {W.}~\bibnamefont
  {Meng}}\ and\ \bibinfo {author} {\bibfnamefont {Z.}~\bibnamefont {Xu}},\
  }\bibfield  {title} {\bibinfo {title} {Quantum speed limits in arbitrary
  phase spaces},\ }\href {https://doi.org/10.48550/arxiv.2210.14278} {\bibfield
   {journal} {\bibinfo  {journal} {arXiv:2210.14278}\ } (\bibinfo {year}
  {2022})}\BibitemShut {NoStop}%
\bibitem [{\citenamefont {Deffner}\ and\ \citenamefont
  {Lutz}(2010)}]{deffner2010}%
  \BibitemOpen
  \bibfield  {author} {\bibinfo {author} {\bibfnamefont {S.}~\bibnamefont
  {Deffner}}\ and\ \bibinfo {author} {\bibfnamefont {E.}~\bibnamefont {Lutz}},\
  }\bibfield  {title} {\bibinfo {title} {Generalized clausius inequality for
  nonequilibrium quantum processes},\ }\href
  {https://doi.org/10.1103/PhysRevLett.105.170402} {\bibfield  {journal}
  {\bibinfo  {journal} {Physical Review Letters}\ }\textbf {\bibinfo {volume}
  {105}},\ \bibinfo {pages} {170402} (\bibinfo {year} {2010})}\BibitemShut
  {NoStop}%
\bibitem [{\citenamefont {Das}\ \emph {et~al.}(2018)\citenamefont {Das},
  \citenamefont {Khatri}, \citenamefont {Siopsis},\ and\ \citenamefont
  {Wilde}}]{das2018}%
  \BibitemOpen
  \bibfield  {author} {\bibinfo {author} {\bibfnamefont {S.}~\bibnamefont
  {Das}}, \bibinfo {author} {\bibfnamefont {S.}~\bibnamefont {Khatri}},
  \bibinfo {author} {\bibfnamefont {G.}~\bibnamefont {Siopsis}},\ and\ \bibinfo
  {author} {\bibfnamefont {M.~M.}\ \bibnamefont {Wilde}},\ }\bibfield  {title}
  {\bibinfo {title} {Fundamental limits on quantum dynamics based on entropy
  change},\ }\href {https://doi.org/https://doi.org/10.1063/1.4997044}
  {\bibfield  {journal} {\bibinfo  {journal} {Journal of Mathematical Physics}\
  }\textbf {\bibinfo {volume} {59}},\ \bibinfo {pages} {012205} (\bibinfo
  {year} {2018})}\BibitemShut {NoStop}%
\bibitem [{\citenamefont {Bekenstein}(1981)}]{bekenstein1981}%
  \BibitemOpen
  \bibfield  {author} {\bibinfo {author} {\bibfnamefont {J.~D.}\ \bibnamefont
  {Bekenstein}},\ }\bibfield  {title} {\bibinfo {title} {Energy cost of
  information transfer},\ }\href {https://doi.org/10.1103/PhysRevLett.46.623}
  {\bibfield  {journal} {\bibinfo  {journal} {Physical Review Letters}\
  }\textbf {\bibinfo {volume} {46}},\ \bibinfo {pages} {623} (\bibinfo {year}
  {1981})}\BibitemShut {NoStop}%
\bibitem [{\citenamefont {Lloyd}(2000)}]{lloyd2000}%
  \BibitemOpen
  \bibfield  {author} {\bibinfo {author} {\bibfnamefont {S.}~\bibnamefont
  {Lloyd}},\ }\bibfield  {title} {\bibinfo {title} {Ultimate physical limits to
  computation},\ }\href {https://doi.org/https://doi.org/10.1038/35023282}
  {\bibfield  {journal} {\bibinfo  {journal} {Nature}\ }\textbf {\bibinfo
  {volume} {406}},\ \bibinfo {pages} {1047} (\bibinfo {year}
  {2000})}\BibitemShut {NoStop}%
\bibitem [{\citenamefont {Lloyd}(2002)}]{lloyd2002}%
  \BibitemOpen
  \bibfield  {author} {\bibinfo {author} {\bibfnamefont {S.}~\bibnamefont
  {Lloyd}},\ }\bibfield  {title} {\bibinfo {title} {Computational capacity of
  the universe},\ }\href {https://doi.org/10.1103/PhysRevLett.88.237901}
  {\bibfield  {journal} {\bibinfo  {journal} {Physical Review Letters}\
  }\textbf {\bibinfo {volume} {88}},\ \bibinfo {pages} {237901} (\bibinfo
  {year} {2002})}\BibitemShut {NoStop}%
\bibitem [{\citenamefont {Ashhab}\ \emph {et~al.}(2012)\citenamefont {Ashhab},
  \citenamefont {de~Groot},\ and\ \citenamefont {Nori}}]{AGN12}%
  \BibitemOpen
  \bibfield  {author} {\bibinfo {author} {\bibfnamefont {S.}~\bibnamefont
  {Ashhab}}, \bibinfo {author} {\bibfnamefont {P.~C.}\ \bibnamefont
  {de~Groot}},\ and\ \bibinfo {author} {\bibfnamefont {F.}~\bibnamefont
  {Nori}},\ }\bibfield  {title} {\bibinfo {title} {Speed limits for quantum
  gates in multiqubit systems},\ }\href
  {https://doi.org/10.1103/PhysRevA.85.052327} {\bibfield  {journal} {\bibinfo
  {journal} {Physical Review A}\ }\textbf {\bibinfo {volume} {85}},\ \bibinfo
  {pages} {052327} (\bibinfo {year} {2012})}\BibitemShut {NoStop}%
\bibitem [{\citenamefont {Caneva}\ \emph {et~al.}(2009)\citenamefont {Caneva},
  \citenamefont {Murphy}, \citenamefont {Calarco}, \citenamefont {Fazio},
  \citenamefont {Montangero}, \citenamefont {Giovannetti},\ and\ \citenamefont
  {Santoro}}]{Caneva2009}%
  \BibitemOpen
  \bibfield  {author} {\bibinfo {author} {\bibfnamefont {T.}~\bibnamefont
  {Caneva}}, \bibinfo {author} {\bibfnamefont {M.}~\bibnamefont {Murphy}},
  \bibinfo {author} {\bibfnamefont {T.}~\bibnamefont {Calarco}}, \bibinfo
  {author} {\bibfnamefont {R.}~\bibnamefont {Fazio}}, \bibinfo {author}
  {\bibfnamefont {S.}~\bibnamefont {Montangero}}, \bibinfo {author}
  {\bibfnamefont {V.}~\bibnamefont {Giovannetti}},\ and\ \bibinfo {author}
  {\bibfnamefont {G.~E.}\ \bibnamefont {Santoro}},\ }\bibfield  {title}
  {\bibinfo {title} {Optimal {C}ontrol at the {Q}uantum {S}peed {L}imit},\
  }\href {https://doi.org/10.1103/PhysRevLett.103.240501} {\bibfield  {journal}
  {\bibinfo  {journal} {Physical review letters}\ }\textbf {\bibinfo {volume}
  {103}},\ \bibinfo {pages} {240501} (\bibinfo {year} {2009})}\BibitemShut
  {NoStop}%
\bibitem [{\citenamefont {Campbell}\ and\ \citenamefont
  {Deffner}(2017)}]{Campbell2017}%
  \BibitemOpen
  \bibfield  {author} {\bibinfo {author} {\bibfnamefont {S.}~\bibnamefont
  {Campbell}}\ and\ \bibinfo {author} {\bibfnamefont {S.}~\bibnamefont
  {Deffner}},\ }\bibfield  {title} {\bibinfo {title} {Trade-{O}ff {B}etween
  {S}peed and {C}ost in {S}hortcuts to {A}diabaticity},\ }\href
  {https://doi.org/10.1103/PhysRevLett.118.100601} {\bibfield  {journal}
  {\bibinfo  {journal} {Physical Review Letters}\ }\textbf {\bibinfo {volume}
  {118}},\ \bibinfo {pages} {100601} (\bibinfo {year} {2017})}\BibitemShut
  {NoStop}%
\bibitem [{\citenamefont {Campbell}\ \emph {et~al.}(2018)\citenamefont
  {Campbell}, \citenamefont {Genoni},\ and\ \citenamefont
  {Deffner}}]{Campbell2018}%
  \BibitemOpen
  \bibfield  {author} {\bibinfo {author} {\bibfnamefont {S.}~\bibnamefont
  {Campbell}}, \bibinfo {author} {\bibfnamefont {M.~G.}\ \bibnamefont
  {Genoni}},\ and\ \bibinfo {author} {\bibfnamefont {S.}~\bibnamefont
  {Deffner}},\ }\bibfield  {title} {\bibinfo {title} {Precision thermometry and
  the quantum speed limit},\ }\href {https://doi.org/10.1088/2058-9565/aaa641}
  {\bibfield  {journal} {\bibinfo  {journal} {Quantum Science and Technology}\
  }\textbf {\bibinfo {volume} {3}},\ \bibinfo {pages} {025002} (\bibinfo {year}
  {2018})}\BibitemShut {NoStop}%
\bibitem [{\citenamefont {Mukhopadhyay}\ \emph {et~al.}(2018)\citenamefont
  {Mukhopadhyay}, \citenamefont {Misra}, \citenamefont {Bhattacharya},\ and\
  \citenamefont {Pati}}]{Mukhopadhyay2018}%
  \BibitemOpen
  \bibfield  {author} {\bibinfo {author} {\bibfnamefont {C.}~\bibnamefont
  {Mukhopadhyay}}, \bibinfo {author} {\bibfnamefont {A.}~\bibnamefont {Misra}},
  \bibinfo {author} {\bibfnamefont {S.}~\bibnamefont {Bhattacharya}},\ and\
  \bibinfo {author} {\bibfnamefont {A.~K.}\ \bibnamefont {Pati}},\ }\bibfield
  {title} {\bibinfo {title} {Quantum speed limit constraints on a nanoscale
  autonomous refrigerator},\ }\href
  {https://doi.org/10.1103/PhysRevE.97.062116} {\bibfield  {journal} {\bibinfo
  {journal} {Physical Review E}\ }\textbf {\bibinfo {volume} {97}},\ \bibinfo
  {pages} {062116} (\bibinfo {year} {2018})}\BibitemShut {NoStop}%
\bibitem [{\citenamefont {You}\ and\ \citenamefont {Gu}(2018)}]{you2018}%
  \BibitemOpen
  \bibfield  {author} {\bibinfo {author} {\bibfnamefont {Y.-Z.}\ \bibnamefont
  {You}}\ and\ \bibinfo {author} {\bibfnamefont {Y.}~\bibnamefont {Gu}},\
  }\bibfield  {title} {\bibinfo {title} {Entanglement features of random
  hamiltonian dynamics},\ }\href {https://doi.org/10.1103/PhysRevB.98.014309}
  {\bibfield  {journal} {\bibinfo  {journal} {Physical Review B}\ }\textbf
  {\bibinfo {volume} {98}},\ \bibinfo {pages} {014309} (\bibinfo {year}
  {2018})}\BibitemShut {NoStop}%
\bibitem [{\citenamefont {Cotler}\ \emph {et~al.}(2017)\citenamefont {Cotler},
  \citenamefont {Hunter-Jones}, \citenamefont {Liu},\ and\ \citenamefont
  {Yoshida}}]{cotler2017}%
  \BibitemOpen
  \bibfield  {author} {\bibinfo {author} {\bibfnamefont {J.}~\bibnamefont
  {Cotler}}, \bibinfo {author} {\bibfnamefont {N.}~\bibnamefont
  {Hunter-Jones}}, \bibinfo {author} {\bibfnamefont {J.}~\bibnamefont {Liu}},\
  and\ \bibinfo {author} {\bibfnamefont {B.}~\bibnamefont {Yoshida}},\
  }\bibfield  {title} {\bibinfo {title} {Chaos, complexity, and random
  matrices},\ }\href {https://doi.org/10.1007/JHEP11(2017)048} {\bibfield
  {journal} {\bibinfo  {journal} {Journal of High Energy Physics}\ }\textbf
  {\bibinfo {volume} {2017}},\ \bibinfo {pages} {48} (\bibinfo {year}
  {2017})}\BibitemShut {NoStop}%
\bibitem [{\citenamefont {Vijay}\ and\ \citenamefont
  {Vishwanath}(2018)}]{vijay2018}%
  \BibitemOpen
  \bibfield  {author} {\bibinfo {author} {\bibfnamefont {S.}~\bibnamefont
  {Vijay}}\ and\ \bibinfo {author} {\bibfnamefont {A.}~\bibnamefont
  {Vishwanath}},\ }\bibfield  {title} {\bibinfo {title} {Finite-temperature
  scrambling of a random hamiltonian},\ }\href
  {https://doi.org/10.48550/arxi.1803.08483} {\bibfield  {journal} {\bibinfo
  {journal} {arXiv:1803.08483}\ } (\bibinfo {year} {2018})}\BibitemShut
  {NoStop}%
\bibitem [{\citenamefont {Yoshida}\ and\ \citenamefont
  {Yao}(2019)}]{yoshida2019}%
  \BibitemOpen
  \bibfield  {author} {\bibinfo {author} {\bibfnamefont {B.}~\bibnamefont
  {Yoshida}}\ and\ \bibinfo {author} {\bibfnamefont {N.~Y.}\ \bibnamefont
  {Yao}},\ }\bibfield  {title} {\bibinfo {title} {Disentangling scrambling and
  decoherence via quantum teleportation},\ }\href
  {https://doi.org/10.1103/PhysRevX.9.011006} {\bibfield  {journal} {\bibinfo
  {journal} {Phys. Rev. X}\ }\textbf {\bibinfo {volume} {9}},\ \bibinfo {pages}
  {011006} (\bibinfo {year} {2019})}\BibitemShut {NoStop}%
\bibitem [{\citenamefont {Gharibyan}\ \emph {et~al.}(2018)\citenamefont
  {Gharibyan}, \citenamefont {Hanada}, \citenamefont {Shenker},\ and\
  \citenamefont {Tezuka}}]{gharibyan2017}%
  \BibitemOpen
  \bibfield  {author} {\bibinfo {author} {\bibfnamefont {H.}~\bibnamefont
  {Gharibyan}}, \bibinfo {author} {\bibfnamefont {M.}~\bibnamefont {Hanada}},
  \bibinfo {author} {\bibfnamefont {S.~H.}\ \bibnamefont {Shenker}},\ and\
  \bibinfo {author} {\bibfnamefont {M.}~\bibnamefont {Tezuka}},\ }\bibfield
  {title} {\bibinfo {title} {Onset of random matrix behavior in scrambling
  systems},\ }\href {https://doi.org/10.1007/JHEP07(2018)124} {\bibfield
  {journal} {\bibinfo  {journal} {Journal of High Energy Physics}\ }\textbf
  {\bibinfo {volume} {2018}},\ \bibinfo {pages} {124} (\bibinfo {year}
  {2018})}\BibitemShut {NoStop}%
\bibitem [{\citenamefont {Nahum}\ \emph {et~al.}(2017)\citenamefont {Nahum},
  \citenamefont {Ruhman}, \citenamefont {Vijay},\ and\ \citenamefont
  {Haah}}]{nahum2017}%
  \BibitemOpen
  \bibfield  {author} {\bibinfo {author} {\bibfnamefont {A.}~\bibnamefont
  {Nahum}}, \bibinfo {author} {\bibfnamefont {J.}~\bibnamefont {Ruhman}},
  \bibinfo {author} {\bibfnamefont {S.}~\bibnamefont {Vijay}},\ and\ \bibinfo
  {author} {\bibfnamefont {J.}~\bibnamefont {Haah}},\ }\bibfield  {title}
  {\bibinfo {title} {Quantum entanglement growth under random unitary
  dynamics},\ }\href {https://doi.org/10.1103/PhysRevX.7.031016} {\bibfield
  {journal} {\bibinfo  {journal} {Physical Review X}\ }\textbf {\bibinfo
  {volume} {7}},\ \bibinfo {pages} {031016} (\bibinfo {year}
  {2017})}\BibitemShut {NoStop}%
\bibitem [{\citenamefont {Hunter-Jones}(2019)}]{nicholas2019}%
  \BibitemOpen
  \bibfield  {author} {\bibinfo {author} {\bibfnamefont {N.}~\bibnamefont
  {Hunter-Jones}},\ }\bibfield  {title} {\bibinfo {title} {Unitary designs from
  statistical mechanics in random quantum circuits},\ }\href
  {https://doi.org/10.48550/arxiv.1905.12053} {\bibfield  {journal} {\bibinfo
  {journal} {arXiv:1905.12053}\ } (\bibinfo {year} {2019})}\BibitemShut
  {NoStop}%
\bibitem [{\citenamefont {Skinner}\ \emph {et~al.}(2019)\citenamefont
  {Skinner}, \citenamefont {Ruhman},\ and\ \citenamefont
  {Nahum}}]{skinner2019}%
  \BibitemOpen
  \bibfield  {author} {\bibinfo {author} {\bibfnamefont {B.}~\bibnamefont
  {Skinner}}, \bibinfo {author} {\bibfnamefont {J.}~\bibnamefont {Ruhman}},\
  and\ \bibinfo {author} {\bibfnamefont {A.}~\bibnamefont {Nahum}},\ }\bibfield
   {title} {\bibinfo {title} {Measurement-induced phase transitions in the
  dynamics of entanglement},\ }\href
  {https://doi.org/10.1103/PhysRevX.9.031009} {\bibfield  {journal} {\bibinfo
  {journal} {Phys. Rev. X}\ }\textbf {\bibinfo {volume} {9}},\ \bibinfo {pages}
  {031009} (\bibinfo {year} {2019})}\BibitemShut {NoStop}%
\end{thebibliography}%

\end{document}